\documentclass[amsmath,amssymb,pra,superscriptaddress]{revtex4}
\usepackage{graphicx}
\usepackage{dcolumn}
\usepackage{bm}
\usepackage{subfigure}
\usepackage{booktabs}
\usepackage{color}
\newcounter{one}
\newcommand{\bra}[1]{\langle #1 |}
\newcommand{\ket}[1]{| #1 \rangle}

\begin{document}

\title{Maximization of thermal entanglement of arbitrarily interacting two qubits}

\author{Tomotaka Kuwahara}
\affiliation{Department of Physics, The University of Tokyo, Komaba, Meguro, Tokyo 153-8505}
\author{Naomichi Hatano}
\affiliation{Institute of Industrial Science, The University of Tokyo, Komaba, Meguro, Tokyo 153-8505}

\begin{abstract}
We investigate the thermal entanglement of interacting two qubits. We maximize it by tuning a local Hamiltonian under a given interaction Hamiltonian. We prove that the optimizing local Hamiltonian takes a simple form which dose not depend on the temperature and that the corresponding optimized thermal entanglement decays as $1/(T\log T)$ at high temperatures. We also find that at low temperatures the thermal entanglement is maximum without any local Hamiltonians and that the second derivative of the maximized thermal entanglement changes discontinuously at the boundary between the high- and low-temperature phases.
\end{abstract}
 
\maketitle

\section{Introduction}
Quantum entanglement plays an essential role in quantum information processing~\cite{Nielsen}. Various kinds of investigation have been carried out to understand properties of entanglement for the last two decades~\cite{horodecki, Amico}.  The thermal entanglement~\cite{Arnesen}, which is entanglement of thermal equilibrium states, is one of the important concepts because it shows us the effect of thermal fluctuations on entanglement. Thermal disturbances generally cause disentanglement and have serious effects on quantum information processing. Therefore, many schemes have been proposed to protect entanglement from thermal disturbances~\cite{Sabrina,Manicini,Romano,Abliz,Plastina,Fardin,Sun,Chan,Gurkan,Kamta}. As one of these schemes, a lot of attention has been paid to methods based on manipulation of local Hamiltonians~\cite{Abliz,Plastina,Fardin,Sun,Chan,Gurkan,Kamta}; for example, in quantum spin systems, bipartite thermal entanglement can be enhanced by modulating external magnetic fields. In the present paper, we focus on a simple question as to how much entanglement can be generated by optimizing the local Hamiltonian. We give a theoretical limit of entanglement enhancement by manipulation of the local Hamiltonians.

Relationships between the thermal entanglement and local parameters have been investigated especially in bipartite quantum spin systems~\cite{Fardin,Sun,Chan,Gurkan,Bose,Arnesen,XWang,GFZhang,Asoudeh,Kamta}. From these researches, behavior of the thermal entanglement under external magnetic fields may be understood in the cases of almost all interactions. However, little has been reported on the \textit{maximization} problem of the thermal entanglement; in the case of the bipartite $XY$~spin model, this problem has been solved only numerically~\cite{Chan}. Until now, there are no analytical approaches to optimizing the thermal entanglement of arbitrarily interacting two qubits.

In the present paper, we will answer the following question: given a system of two qubits which interact via an arbitrary interaction Hamiltonian, how can we maximize the thermal entanglement between these two qubits by changing only the local Hamiltonian? A naive approach to this problem may be to solve the optimization problem numerically. However, this problem has six local parameters in total and the functional forms of entanglement measures such as the concurrence~\cite{Wootter} and the negativity~\cite{Vidal} are very complicated. Thus, for an arbitrary interaction, it is difficult to solve this optimization problem numerically. Therefore, we employ perturbation techniques and utilize symmetric properties in order to determine the optimizing local Hamiltonian analytically. In this way, for all kinds of interaction, we give general properties of the optimized entanglement.

Our main results are the following:
\begin{enumerate}
\item{} We find that at low temperatures the thermal entanglement is maximum without any local Hamiltonians, whereas at high temperatures it is maximized by non-zero local fields.
We refer to the former temperature range as the low-temperature phase and the latter temperature range as the high-temperature phase.
The secondary differentiation of the maximized entanglement is discontinuous at the phase boundary.
\item{} In the high-temperature phase, the functional form of the optimizing local Hamiltonian is independent of the temperature; only the coefficients depend on the temperature.
\item{} The optimized entanglement, enhanced by a local Hamiltonian in the high-temperature phase, decreases with increasing temperature as $1/(T \log T)$.
\item{}If the interaction Hamiltonian has no degeneracy of its eigenvalues, the entanglement is maximized without local Hamiltonians over a finite range of the low-temperature phase.
\item{}If the interaction Hamiltonian has degeneracy, the low-temperature phase shrinks to the zero-temperature point. The optimizing local Hamiltonian becomes infinitesimal and the optimized entanglement becomes full in the low-temperature limit.
\end{enumerate}
Our paper is organized as follows. In Section~II, we state the main problem after symmetry consideration. In Section~III, we give the main theorems on the entanglement optimization. In Section~IV, we show numerical results of the optimizing local parameters, the boundary temperatures and the singularity at the phase boundary. We also argue that the two phases appear because of competition between the purifying effect and the decoupling effect both of the local Hamiltonian. Finally, in Section~V, a discussion concludes the paper. 
%In Appendix~A, we discuss the convergence of the approximation obtained in Section~3. In Appendix~B, we complement the proof of Lemma~2 in Section~3. In Appendix~C, we prove the equation obtained in 

\section{Entanglement optimization problem}
First, we set the fundamental framework of the present problem.
We consider a ${2 \otimes 2}$ system of $\sigma_1$ and  $\sigma_2$. %we note that it is equivalent to the $\frac{1}{2}$spin-system
The most general form of the Hamiltonian of this system is given as follows:
\begin{align}
H_{\textrm{tot}} &\equiv H_{\textrm{int}} + H_{\textrm{LO}}, \notag \\
H_{\textrm{int}}  &\equiv \sum_{i,j = x,y,z}J_{ij} \sigma_1^i\otimes \sigma_2^j  ,\notag \\
H_{\textrm{LO}}  &\equiv \sum_{i = x,y,z} ( h_1^i \sigma_1^i\otimes I +h_2^i I \otimes \sigma_2^i), \label{Hamiltonians}
\end{align}
where ${\{\sigma_1^i\}_{i=x,y,z}}$ and ${\{\sigma_2^i\}_{i=x,y,z}}$ are the Pauli matrices, $H_{\textrm{int}}$ is an interaction Hamiltonian, and $H_{\textrm{LO}}$ is a local Hamiltonian. We assume that $\{J_{ij}\}_{i, j =x, y, z}$ are fixed and independent of the temperature, whereas we can change the parameters ${\{h_1^x,h_1^y,h_1^z,h_2^x,h_2^y,h_2^z\}}$ arbitrarily.

We parametrize the local fields in the polar coordinates as 
\begin{align}
\{h_1^i\}_{i=x,y,z} &=\{h_1 \sin \theta_1 \cos \phi_1 ,h_1 \sin\theta_1\sin \phi_1,h_1\cos \theta_1\}  ,\notag \\
\{h_2^i\}_{i=x,y,z} &=\{h_2\sin \theta_2 \cos \phi_2 ,h_2 \sin\theta_2\sin \phi_2,h_2 \cos \theta_2\}. \notag \\
\end{align}
Hereafter, we use the parametrization 
\begin{align}
h\equiv \frac{h_1+h_2}{2}, \ \ \zeta \equiv \frac{h_1-h_2}{h_1+h_2}, 
\end{align}
where ${-1\le\zeta\le1}$ and ${h\ge0}$; in other words, 
\begin{align}
h_1=(1+\zeta)h, \ \  h_2=(1-\zeta)h.
\end{align}
Then, the four eigenvalues of $H_{\textrm{LO}}$ are 
\begin{align}
\{-2h,-2\zeta h,2\zeta h,2h\},
\end{align}
where we define the corresponding eigenstates as ${\{\ket{{--}},\ket{{-+}},\ket{{+-}},\ket{{++}}\}}$.

The density matrix in thermal equilibrium is 
\begin{align}
\rho &= \frac{e^{-\beta H_{\textrm{tot}}}}{Z},     \label{density matrix}
\end{align}
where ${Z =\textrm{tr}( e^{-\beta H_{\textrm{tot}}}) }$ is the partition function and ${\beta = 1/k T}$ with $k$ the Boltzmann constant. In order to quantify entanglement, we adopt the negativity~\cite{Vidal} as an entanglement measure. The negativity is defined as the trace norm of a partially transposed density matrix:
\begin{align} 
N(\rho) &\equiv || \rho^{T_1} ||_1 -1  \notag \\
                  &=\textrm{max}(-2\lambda_-,0) ,\label{negativity}
\end{align} 
where $||\ ||_1$ is the trace norm, $T_1$ denotes the transpose with respect to only $\sigma_1$, and $\lambda_-$ is the minimum, possibly negative eigenvalue of $\rho^{T_1}$. The second equation of (\ref{negativity}) comes from the fact that  $\rho^{T_1}$ can have only one negative eigenvalue, if any~\cite{Augusiak}. Thus, the present entanglement optimization problem is equivalent to finding the values of ${\{h_1^x,h_1^y,h_1^z,h_2^x,h_2^y,h_2^z\}}$ which maximize $N(\rho)$ for an arbitrary fixed interaction $H_{\textrm{int}}$. 

Before presenting our main results on the entanglement optimization,
we prove the following Lemma~1 to simplify the present entanglement optimization problem. 

\textit{Lemma~1}. 
By local unitary transformations of $H_{\textrm{int}}$, we can eliminate the interaction parameters $\{J_{ij}\}_{i\neq j}$ and reduce it to the form
\begin{align}
H_{\textrm{int}} = \sum_{i= x,y,z} J_{i} \sigma_1^i\otimes \sigma_2^i . \label{diagform}
\end{align}
We can also choose the parameters $\{J_{x},J_{y},J_{z}\}$ such that ${\{J_{x},J_{y}\}\ge J_{z}\ge0}$ or ${0\ge J_{z} \ge \{J_{x},J_{y}\}}$.

In spin-1/2 systems, this means that we can transform any interactions including the Dzyaloshinskii-Moriya (DM)~\cite{Dzyaloshinskii,Moriya,Fardin} interaction into a ferromagnetic or an anti-ferromagnetic Heisenberg exchange interaction.

\textit{Proof}. 
We can prove this Lemma by applying a singular value decomposition~\cite{Matrix Analysis} to the matrix $(\hat{J})_{ij}\equiv J_{ij}$. In this case, the singular value decomposition ${\hat{U} \hat{J} \hat{W}}$ is performed by $3\times3$ real orthogonal transformations $\hat{U}$ and $\hat{W}$ of the three-dimensional spin spaces of the spins~1 and 2, respectively. A real orthogonal transformation is composed of rotation and inversion operations, but inversion operations cannot be performed by unitary transformations. Therefore, we remove the inversion operations from the real orthogonal transformation of the singular value decomposition and restrict ourselves only to the rotation operations, which means ${\det \hat{U} = \det \hat{W} = 1}$.  In other words, we rotate ${\vec{\sigma}_1=\{\sigma_{1}^x,\sigma_{1}^y,\sigma_{1}^z\}}$ with $\hat{U}$ and ${\vec{\sigma}_2=\{\sigma_{2}^x,\sigma_{2}^y,\sigma_{2}^z\}}$ with $\hat{W}$. Then we can transform $\{J_{ij}\}_{i,j=x,y,z}$ into the antiferromagnetic cases ${\{J_{x},J_{y}\}\ge J_{z}\ge0}$ or the ferromagnetic cases ${0\ge J_{z} \ge \{J_{x},J_{y}\}}$, with the other elements ${\{J_{ij}\}_{i\neq j}}$ put to zero. Here, we choose the $z$-axis so that $|J_{z} |$ is the least of $\{|J_{i} |\}_{i=x,y,z}$. Thus, Lemma~1 is proved.

Let us show an example in the case of the $XXZ$~model with the $z$-component of the DM interaction. The Hamiltonian of such a system is given by
\begin{align}
H_{\textrm{int}}  \equiv& J \sigma_1^x\otimes \sigma_2^x + J \sigma_1^y\otimes \sigma_2^y + J_z \sigma_1^z\otimes \sigma_2^z  \notag \\
                       & + D_z(\sigma_1^x\otimes \sigma_2^y -  \sigma_1^y\otimes \sigma_2^x),
\end{align}
where $J$ and $J_z$ are the real coupling coefficients and $D_z$ is the $z$-component of the DM interaction. In the case of ${J=1}$, ${J_z=-2}$ and ${D_z=1}$, we can transform $\{J_x, J_y, J_z, D_z\}$ into $\{-\sqrt{2}, -\sqrt{2}, -2,0\}$ by rotating the spin 1 by $135$ degrees around the $z$-axis, namely into
\begin{align}
H_{\textrm{int}}  =-\sqrt{2}\sigma_1^x\otimes \sigma_2^x -\sqrt{2}\sigma_1^y\otimes \sigma_2^y -2 \sigma_1^z\otimes \sigma_2^z. 
\end{align}
This is an antiferromagnetic Heisenberg interaction.
To attain this result, first, the singular value decomposition transforms $\{J_x,J_y,J_z,D_z\}$ into $\{\sqrt{2},\sqrt{2}, 2,0\}$ by rotating the spin 1
by $-45$ degrees around the $z$-axis and inverting the $z$-axis of the spin. 
Next, we remove the inversion of the $z$-axis because it cannot be performed by unitary operations, and thereby transform $\{J_x, J_y, J_z, D_z\}$ into $\{\sqrt{2}, \sqrt{2}, -2,0\}$. By changing the rotation angle from $-45$ to $135$, we can invert the signs of $J_x$ and $J_y$ and arrive at $\{J_x, J_y, J_z, D_z\}=\{-\sqrt{2}, -\sqrt{2}, -2,0\}$.

In the following, based on Lemma 1, we always use the diagonalized form~(\ref{diagform}) of the interaction parameters with ${\{J_{x},J_{y}\}\ge J_{z}\ge0}$ or ${0\ge J_{z} \ge \{J_{x},J_{y}\}}$.
We now have all the necessary ingredients to state the main theorems.

 \section{Main analytical results}
 In the present section, we analytically discuss the optimization problem. The main conclusion of the present section is that the negativity is maximized by the parameters $\{h_1^x,h_1^y,h_1^z,h_2^x,h_2^y,h_2^z\} = \{0,0, h_{\textrm{op}},0,0,-h_{\textrm{op}}\}$. The optimizing parameter $h_{\textrm{op}}$ must be very large at high temperatures, whereas it may be $0$ at low temperatures.  
 
 \subsection{Optimization in the high-temperature limit}
Let us first discuss the optimization problem in the high-temperature limit. 

\textit{Theorem~1}.
 In the high-temperature limit ${\beta\rightarrow0}$, the local parameters which maximize the entanglement $N(\rho)$ are given in the form of $\{h_1^x,h_1^y,h_1^z,h_2^x,h_2^y,h_2^z\} = \{0,0, h_{\textrm{op}},0,0,-h_{\textrm{op}}\}$. The optimizing value $h_{\textrm{op}}$ is given by the solution of the following equation:
 \begin{align}
e^{2h'_{\textrm{op}}}& \simeq \frac{8h'^2_{\textrm{op}}}{\beta |J_x+J_y|} \quad \mbox{as} \quad \beta \rightarrow 0,  \label{happro} 
\end{align}
where
\begin{align}
h'_{\textrm{op}}&\equiv \beta h_{\textrm{op}},
\end{align}
and the optimized entanglement $N_{\textrm{op}}$ asymptotically behaves as
\begin{align}
N_{\textrm{op}}(\rho) &\simeq \beta \frac{|J_x+J_y|}{2h'_{\textrm{op}}} - 2e^{-2 h'_{\textrm{op}}} \notag \\
 &\simeq \beta \frac{|J_x+J_y|}{2h'_{\textrm{op}}}  \Bigl(1- \frac{1}{2h'_{\textrm{op}}} \Bigr) \quad \mbox{as} \quad \beta \rightarrow 0,               \label{Nappro}
\end{align}
where we used Eq.~(\ref{happro}) upon moving from the first line to the second line.

The leading order of the solution of {Eq.~(\ref{happro})} is given by 
\begin{align}
h'_{\textrm{op}} \simeq \frac{1}{2} \log \frac{1}{\beta} + \frac{1}{2} \log \frac{8}{ |J_x+J_y|} . \label{hopappro}
\end{align}
We can thereby obtain the following simpler asymptotes: 
\begin{align}
h_{\textrm{op}}&\simeq \frac{\log 1/\beta}{2\beta} \quad \mbox{as} \quad \beta \rightarrow 0, \label{theorem1} \\
N_{\textrm{op}}(\rho) &\simeq \beta \frac{|J_x+J_y|}{\log 1/\beta} \quad \mbox{as} \quad \beta \rightarrow 0. \label{theorem1N}
\end{align}
That is, the optimizing value $h_{\textrm{op}}$ depends only on the temperature and the optimized negativity decays in the form ${1/(T\log T)}$ in the limit ${\beta\rightarrow0}$. In Appendix~A, we compare the asymptotes of {Eqs.~(\ref{happro})} and (\ref{Nappro}) with those of {Eqs.~(\ref{theorem1})} and (\ref{theorem1N})

\textit{Proof}. 
We prove Theorem~1 in the following steps. First, we prove in Lemma~2 that the optimizing local parameter ${h_{\textrm{op}}}$ is greater than or equal to ${(\log1/\beta)/(2\beta)}$ in the high-temperature limit and the optimized thermal state is nearly a pure state. The entanglement of the state comes from perturbations to the pure state. Then, we calculate the negativity approximately by perturbation method in Lemma~4. Using this expression, we finally solve the maximization problem for each local parameter. 

First, we determine a lower bound of the optimizing value $h_{\textrm{op}}$ and prove that the optimized thermal state is a nearly pure state.
For this purpose, we prove the following Lemma~2.

\textit{Lemma~2}. 
A necessary condition for the existence of the entanglement in the high-temperature limit under a fixed interaction Hamiltonian $H_{\textrm{int}}$ is given by
\begin{align}
\beta h > \frac{\log 1/\beta}{2}\quad \mbox{as} \quad \beta \rightarrow 0. \label{Lemma2}
\end{align}
This Lemma~2 shows that ${(\log1/\beta)/(2\beta)}$ is a lower bound of the optimizing value of $h_{\textrm{op}}$.

\textit{Proof}. 
We firstly prove that we need a non-zero value of $\beta h$ for the existence of the entanglement in the high-temperature limit ${\beta\rightarrow0}$. 
In other words, we need $h$ at least of order $1/\beta$.
In order to show this, we consider a general necessary condition for the existence of the entanglement given by~\cite{Verstraete1}
\begin{align}
\lambda_1  \ge  \lambda_3+2 \sqrt{\lambda_2 \lambda_4} \ge 3 \lambda_4 , \label{necessary}
\end{align} 
where $\{ \lambda_\mu \}_{\mu=1}^4$ are the eigenvalues of the density matrix $\rho$ in the non-ascending order (${\lambda_1\ge  \lambda_2\ge \lambda_3\ge \lambda_4}$). 
Let us define the eigenvalues of ${H_{\textrm{tot}} =H_{\textrm{LO}}+H_{\textrm{int}}}$ as $\{E_\mu \}_{\mu=1}^4$ in the non-descending order (${E_1\le  E_2\le E_3\le E_4}$). Equation~(\ref{density matrix}) gives the eigenvalues of $\rho$ as $\{e^{-\beta E_\mu}/Z\}_{\mu=1}^4$, and therefore the inequality~(\ref{necessary}), or $e^{-\beta E_1} \ge 3 e^{-\beta E_4}$, gives
\begin{align}
\beta (E_4 -E_1) \ge \log3. \label{necessary2}
\end{align}
Here, $H_{\textrm{int}}$ is a constant matrix and hence ${\beta H_{\textrm{int}}\rightarrow 0}$ as ${\beta \rightarrow 0}$.
If we let ${H_{\textrm{LO}}}$ be of the same order as ${H_{\textrm{int}}}$, the left-hand side of (\ref{necessary2}) would vanish in the limit ${\beta \rightarrow 0}$ and (\ref{necessary}) would not be satisfied. Therefore, we have to make ${H_{\textrm{LO}}}$ much greater than ${H_{\textrm{int}}}$, and then the eigenvalues of ${H_{\textrm{tot}}}$ should converge to those of ${H_{\textrm{LO}}}$, ${\{-2 h,-2 \zeta h,2 \zeta h,2h\}}$ in the limit $\beta \rightarrow 0$. With ${E_1 \rightarrow -2h}$ and ${E_4 \rightarrow 2h}$, the inequality (\ref{necessary2}) reduces to the following inequality:
\begin{align}
\beta h \ge \frac{\log3}{4}.  \label{betahc}
\end{align}
This inequality means that  we need a non-zero value of $\beta h$ in the high-temperature limit ${\beta\rightarrow0}$.
In other words, we need to make $h$ grow as $1/\beta$ at least, in order for the entanglement to exist in the limit ${\beta\rightarrow0}$.

Next, we derive an approximation of the density matrix, and then obtain Eq.~(\ref{Lemma2}) by utilizing the Peres-Horodecki criterion~\cite{Peres,Horodecki}, which is a necessary and sufficient condition for the existence of the entanglement.
In the present optimization problem, we fix $H_{\textrm{int}}$ to a constant matrix, and therefore we have ${{\beta H_{\textrm{int}}\rightarrow 0}}$ in the high-temperature limit. We thereby work in the first-order approximation with respect to ${\beta H_{\textrm{int}}}$:
 \begin{align}
Z\rho&=e^{ - H'_{\textrm{LO}} - \beta H_{\textrm{int}}}  \notag \\
&\simeq e^{-H'_{\textrm{LO}}}  - \beta \int_0^1 e^{- (1-x) H'_{\textrm{LO}}} H_{\textrm{int}} e^{-x H'_{\textrm{LO}}} dx  \notag \\
&=e^{-H'_{\textrm{LO}}}  - \beta \sum_{\mu,\nu} f_{\mu\nu} \langle \mu | H_{\textrm{int}} | \nu \rangle \ket{\mu} \bra{\nu} ,\label{perturbation}
\end{align}
where ${Z=\textrm{tr}( e^{ - H'_{\textrm{LO}} - \beta H_{\textrm{int}}})}$, and we let ${H'_{\textrm{LO}}=\beta H_{\textrm{LO}}}$ with ${h' = \beta h}$ as well as
\begin{align}
         f_{\mu\nu} &= \begin{cases}
                           \frac{e^{-E'_\nu}-e^{-E'_\mu}}{E'_\mu - E'_\nu},  &\textrm{for}\ E'_\mu \neq E'_\nu,  \\
                           e^{-E'_\mu}, &  \textrm{for}\ E'_\mu=E'_\nu .
                            \end{cases}  \label{fij}
\end{align}
Here, $\{E'_\mu\}_{\mu=1}^4$ are the eigenvalues of ${H'_{\textrm{LO}}= \beta H_{\textrm{LO}}}$, ${\{-2h',-2\zeta h',2\zeta h',2h'\}}$, and $\{\ket{\mu}\}_{\mu=1}^4$ are the corresponding eigenstates, ${\{\ket{{--}},\ket{{-+}},\ket{{+-}},\ket{{++}}\}}$. 

We then utilize the necessary and sufficient condition for the existence of the entanglement, ${\det \rho^{T_1}<0}$. This has been proved~\cite{Augusiak} to be equivalent to the Peres-Horodecki criterion~\cite{Peres,Horodecki}. 
In the following discussion, among the various terms of the expansion of $\det \rho^{T_1}$, we compare the values of the products including off-diagonal elements (POD) with that of the product of the diagonal elements (PD), which has a positive value. Then a necessary condition for ${\det \rho^{T_1}<0}$ is that POD is greater than or of the same order as the PD.
%Among the various terms of the expansion of ${\det \rho^{T_1}}$, the PD of ${Z\rho^{T_1}}$ is $O(1)$ and the PODs are roughly ${O(e^{h'} \beta^2)}$. Therefore, at least, it is necessary for ${\det \rho^{T_1}<0}$ that $h'$ is ${O(\log 1/\beta)}$.

To analyze the order of the PD and the PODs, we express $\rho^{T_1}$ in the basis $\{\ket{\mu}\}_{\mu=1}^4$ and focus on the main terms for ${\zeta \neq 0}$ and ${\zeta \neq \pm1}$:
\begin{align}
&Z\rho^{T_1}  \notag \\
&\xrightarrow{\beta \rightarrow 0} \begin{pmatrix}  e^{2h'} & a_{12}\frac{\beta e^{2h'}}{h'} & a_{31}\frac{\beta e^{2h'}}{h'} & a_{32}\frac{\beta e^{2\zeta h'}}{h'} \\ a_{21}\frac{\beta e^{2h'}}{h'}  & e^{2\zeta h'} &a_{41}\frac{\beta e^{2h'}}{h'} &a_{42} \frac{\beta e^{2\zeta h'}}{h'} \\ a_{13}\frac{\beta e^{2h'}}{h'}  &a_{14}\frac{\beta e^{2h'}}{h'} &  e^{-2\zeta h'} &  a_{34}\frac{\beta e^{-2\zeta h'}}{h'} \\ a_{23}\frac{\beta e^{2\zeta h'}}{h'}& a_{24} \frac{\beta e^{2\zeta h'}}{h'} &a_{43}\frac{\beta e^{-2\zeta h'}}{h'} & e^{-2 h'}  \end{pmatrix}, \label{Zrho}
\end{align}
where $\{a_{ij}\}$ are determined from Eqs.~(\ref{perturbation}) and (\ref{fij}) and are constants of order 1. Note that on the diagonal of Eq.~\eqref{Zrho}, the second term in Eq.~(\ref{perturbation}) is neglected in comparison to the first term.
Then we compare the orders of the PODs with that of the PD.
The PD is given by ${e^{2h'}  e^{2\zeta h'} e^{-2\zeta h'} e^{-2 h'} =1}$, whereas each POD includes at least two off-diagonal elements.
The maximum of the absolute value of the PODs is of order ${e^{4h'} \beta^2/h'^2}$, which comes from the product ${-e^{2 h'} \times  a_{41}\frac{\beta e^{2h'}}{h'} \times  a_{14} \frac{\beta e^{2h'}}{h'}\times  e^{-2h'}}$. Therefore, it is necessary for ${\det \rho^{T_1}<0}$ that ${e^{4h'} \beta^2/h'^2}$ is greater or of order 1, which is the order of PD. By taking the logarithm of ${e^{4h'} \beta^2/h'^2}$, we can obtain the following inequality as a necessary condition:
\begin{align}
\beta h =h' \ge& \frac{\log 1/\beta}{2} + \frac{\log h'}{2}  \notag \\
                       \ge& \frac{\log 1/\beta}{2} + \frac{1}{2} \log \biggl (\frac{\log 1/\beta}{2} + \frac{\log h'}{2} \biggr) \notag \\
                       \ge& \frac{\log 1/\beta}{2} + \frac{1}{2} \log \biggl (\frac{\log 1/\beta}{2} + \frac{1}{2}\log \Bigl( \frac{\log3}{4} \Bigr) \biggr) \notag \\
                          >& \frac{\log 1/\beta}{2} ,    \label{hlower}
\end{align}
where we utilized (\ref{betahc}) in deriving the third inequality and used the fact ${\beta \rightarrow 0}$ in deriving the last inequality.
Thus, Lemma~2 is proved for ${\zeta \neq 0}$ and ${\zeta \neq \pm1}$. 
For ${\zeta = 0}$ or ${\zeta = \pm1}$, some of the eigenvalues of $H'_{\textrm{LO}}$ are degenerate, which means that $E'_{\mu}$ can be equal to $E'_{\nu}$ in Eq.~(\ref{fij}), and $Z \rho^{T_1}$ is not of the same form as that of Eq.~(\ref{Zrho}). However, the inequality~(\ref{hlower}) still holds as is proved in Appendix~B.

We now consider the negativity (\ref{negativity}) in the range given by (\ref{Lemma2}). We first show in the following Lemma~3 that the optimized negativity in the cases of ${\zeta = \pm1}$ is not large enough.

\textit{Lemma~3}. In the cases of ${\zeta = \pm1}$, the optimized negativity satisfies the following:
\begin{align}
\frac{N_{\textrm{op}}(\rho,\zeta = \pm1)}{\beta} \xrightarrow{\beta \rightarrow 0} 0. \label{degenerate}
\end{align} 
This lemma shows that the optimized negativity in the cases of ${\zeta = \pm1}$ is of a higher order of $\beta$. 
Indeed, we numerically confirmed in the cases of ${\zeta =  \pm1}$ that the entanglement exists, but its amplitude is of order $\beta^2$.

\textit{Proof}.
Let us prove Eq.~(\ref{degenerate}) in the case of ${\zeta =  1}$. The proof for ${\zeta =  -1}$ is almost the same.
We start from the main term of $Z\rho^{T_1}$ for ${\zeta = 1}$ in the representation in the basis $\{\ket{\mu}\}_{\mu=1}^4$: 
\begin{align}
&Z\rho^{T_1}  \notag \\
&\xrightarrow{\beta \rightarrow 0} \begin{pmatrix}  e^{2h'} & a_{12} \beta e^{2h'}  & a_{31}\frac{\beta e^{2h'}}{h'}  &a_{32}\frac{\beta e^{2h'}}{h'} \\ a_{21} \beta e^{2h'}  & e^{2  h'} &a_{41}\frac{\beta e^{2 h'}}{h'}& a_{42} \frac{\beta e^{2  h'}}{h'} \\ a_{13}\frac{\beta e^{2h'}}{h'} & a_{14}\frac{\beta e^{2  h'}}{h'} &  e^{-2  h'} &  a_{34} \beta e^{-2 h'}  \\ a_{23}\frac{\beta e^{2h'}}{h'} &a_{24} \frac{\beta e^{2 h'}}{h'} &a_{43} \beta e^{-2 h'}  & e^{-2 h'}  \end{pmatrix},  \label{zeta1}
\end{align} 
where we used the fact that at $\zeta = 1$ the eigenvalues of $H'_{\textrm{LO}}$ in Eq.~(\ref{fij}) are degenerate as $\{E'_1,E'_2,E'_3,E'_4\}=\{2h',2h',-2h',-2h'\}$.

In order to optimize the negativity, we necessarily consider the region ${h'=\beta h > (\log 1/\beta)/2}$ as is given in Lemma~2. Therefore, we can use the fact ${e^{2h'} > \beta^{-1}}$ in (\ref{zeta1}).
Of the elements of the matrix (\ref{zeta1}), the $(1,1)$ and $(2,2)$ elements are of order $\beta^{-1}$ or greater, whereas the $(3,3)$, $(4,4)$, $(3,4)$ and $(4,3)$ elements are of order $\beta^{1}$ or less. The other elements are approximately of order $1$. We therefore break up the matrix (\ref{zeta1}) in the form
\begin{align}
\rho^{T_1}  &= \frac{1}{Z}  \begin{pmatrix}  e^{2h'} & 0  & 0 &0 \\ 0 & e^{2  h'} &0&0 \\ 0& 0 &  0 &  0 \\ 0 & 0   & 0 &0  \end{pmatrix}                     \notag \\
 +&\frac{1}{Z}  \begin{pmatrix}  0 & a_{12} \beta e^{2h'}  & a_{31}\frac{\beta e^{2h'}}{h'}  &a_{32}\frac{\beta e^{2h'}}{h'} \\ a_{21} \beta e^{2h'}  & 0&a_{41}\frac{\beta e^{2 h'}}{h'}& a_{42} \frac{\beta e^{2  h'}}{h'} \\ a_{13}\frac{\beta e^{2h'}}{h'} & a_{14}\frac{\beta e^{2  h'}}{h'} &  0 &  0 \\ a_{23}\frac{\beta e^{2h'}}{h'} &a_{24} \frac{\beta e^{2 h'}}{h'}   & 0 &0  \end{pmatrix} + O(\beta^2), \label{rhoT1}
\end{align}
where ${Z\simeq 2e^{2h'} + 2e^{-2h'} \simeq 2/\beta}$, and therefore the first term is the dominant term of order $1$, whereas the second term is of order $\beta^1$. 
The eigenvalues of the dominant term are ${\{e^{2h'}/Z, e^{2h'}/Z,0,0\}}$ and the corresponding eigenstates are ${\{\ket{{--}},\ket{{-+}},\ket{{+-}},\ket{{++}}\}}$.
A negative eigenvalue can appear when the degeneracy of the two zero eigenvalues of the states ${\ket{{+-}}}$ and ${\ket{{++}}}$ is resolved by perturbation.
Then, the level repulsion between them makes one of them positive and the other negative.
However, the first-order perturbation of the second term of Eq.~\eqref{rhoT1} dose not resolve the degeneracy of the zero eigenvalues.
Therefore, the negative eigenvalue must be produced in a higher order of $\beta$ in the case of $\zeta=1$.
Thus, Lemma~3 is proved. 
 We focus on the cases ${\zeta \neq \pm1}$ hereafter.

Using the lower bound (\ref{Lemma2}) of the optimizing parameter $h_{\textrm{op}}$, we next prove that the optimized thermal state is a nearly pure state in the cases of ${\zeta \neq \pm1}$.
For this purpose, we consider the eigenstates of the perturbed density matrix. We define the perturbed eigenstates of ${\beta H_{\textrm{tot}}= H'_{\textrm{LO}}+\beta H_{\textrm{int}}}$ as ${\{\ket{{--'}},\ket{{-+'}},\ket{{+-'}},\ket{{++'}}\}}$ corresponding to the eigenstates ${\{\ket{{--}},\ket{{-+}},\ket{{+-}},\ket{{++}}\}}$ of $H'_{\textrm{LO}}$, respectively, and their eigenvalues as ${\{2h'-\beta \delta \epsilon_1, 2\zeta h'-\beta \delta \epsilon_2,-2\zeta h'-\beta \delta \epsilon_3, -2h'-\beta \delta \epsilon_4 \}}$, where ${\{\delta \epsilon_i\}_{i=1}^4}$ are the perturbative changes due to $H_{\textrm{int}}$, which are of order $1$.
Then the density matrix is given by the summation over these four states. 
In the high-temperature limit ${\beta \rightarrow 0}$, the mixing ratio ${\{\lambda_{--'}, \lambda_{-+'},\lambda_ {+-'},\lambda_ {++'}\}}$ of the states ${\{\ket{{--'}},\ket{{-+'}},\ket{{+-'}},\ket{{++'}}\}}$ are 
\begin{align}
&\{\lambda_{--'}, \lambda_{-+'},\lambda_ {+-'},\lambda_ {++'}\} \notag \\
=&\frac{1}{Z} \{e^{2h'-\beta \delta \epsilon_1}, e^{2\zeta h'-\beta \delta \epsilon_2},e^{-2\zeta h'-\beta \delta \epsilon_3}, e^{-2h'-\beta \delta \epsilon_4} \},\label{eigenva}
\end{align}
where
\begin{align}
Z=e^{2h'-\beta \delta \epsilon_1}+e^{2\zeta h'-\beta \delta \epsilon_2}+e^{-2\zeta h'-\beta \delta \epsilon_3}+e^{-2h'-\beta \delta \epsilon_4}.\label{eigenvaZ}
\end{align}
In the region ${h'>(\log 1/\beta)/2}$, which is the lower bound of $h'_{\textrm{op}}$, we have
\begin{align}
&\frac{\lambda_{-+'}}{\lambda_{--'}} = e^{-2(1-\zeta)h'+\beta \delta \epsilon_1 -\beta \delta \epsilon_2} < \beta^{1-\zeta}  e^{-\beta \delta \epsilon_1 +\beta \delta  \epsilon_2},  \notag \\
&\frac{\lambda_{+-'}}{\lambda_{--'}} = e^{-2(1+\zeta)h'+\beta \delta \epsilon_1-\beta \delta \epsilon_3} < \beta^{1+\zeta} e^{-\beta \delta \epsilon_1 +\beta \delta  \epsilon_3} ,\notag \\
&\frac{\lambda_{++'}}{\lambda_{--'}} = e^{-2h'+\beta \delta \epsilon_1 -\beta \delta \epsilon_4} < \beta^{2}  e^{-\beta \delta \epsilon_1 +\beta \delta  \epsilon_4}.  
\end{align}
Since the right-hand sides of the inequalities vanish in the limit ${\beta \rightarrow 0}$, we deduce that the optimized thermal state is a nearly pure state of $\ket{{--'}}$ in the high-temperature limit ${\beta \rightarrow 0}$ when ${\zeta \neq \pm1}$.

Next, we perturbatively calculate the negativity in the cases of ${\zeta \neq \pm1}$. Since the optimized state is a nearly pure state of $\ket{{--'}}$, we regard the other contributions ${\{\ket{{-+'}},\ket{{+-'}},\ket{{++'}}\}}$ as perturbation:
\begin{align}
\rho_0 &=\ket{{--'}}\bra{{--'}}, \notag \\
\delta \rho &= \sum_{\{i={-+'}, {+-'}, {++'}\}} \lambda_i \rho_i,  \label{deltarho}
\end{align}
where ${\{\ket{{--'}},\ket{{-+'}},\ket{{+-'}},\ket{{++'}}\}}$ are the eigenstates of ${\beta H_{\textrm{tot}}= H'_{\textrm{LO}}+\beta H_{\textrm{int}}}$ as has been stated.
In order to calculate the negativity approximately, we derive the expression for the perturbation of the negativity caused by an infinitesimal variation of the density matrix.
  
\textit{Lemma~4}. When the negativity has a non-zero value, the first-order perturbation of the negativity is given by 
\begin{align}
N(\rho_0+\delta \rho) \simeq N(\rho_0) - 2\langle \phi_-| \delta \rho^{T_1} | \phi_- \rangle, \label{negaperturbation}
\end{align} 
where we refer to the eigenstate corresponding to the negative eigenvalue of $\rho_0^{T_1}$ as $\ket{\phi_-}$. 

\textit{Proof}.
The non-zero negativity is given by the negative eigenvalue $\lambda_{-}$ of the partial transpose of the density matrix, $\rho_0^{T_1}$, as is defined in (\ref{negativity}). Because of the linearity of the partial transpose, if $\rho_0$ changes into ${\rho_0 + \delta \rho}$, $\rho_0^{T_1}$ also changes into ${\rho_0^{T_1} + \delta \rho^{T_1}}$. Moreover, the eigenstate of $\rho_0^{T_1}$ corresponding to $\lambda_{-}$ is not degenerate because $\lambda_{-}$ is the only possible negative eigenvalue~\cite{Augusiak}. Then, from the general perturbation theory for $\lambda_{-}$, we have Eq.~(\ref{negaperturbation}) in the first order.

From Eqs.~(\ref{deltarho}) and (\ref{negaperturbation}), we can calculate the negativity in the present case of ${\zeta \neq \pm1}$ as
\begin{align}
&N\biggl(\sum_{i=1}^4 \lambda_i \rho_i \biggr)  \notag \\
=& N(\ket{{--'}}) - \sum_{\{i={-+'}, {+-'}, {++'}\}}\Bigl( 2\lambda_i \langle \phi_-|  \rho_i^{T_1} | \phi_- \rangle + O(\lambda_i^2) \Bigr).\label{negativityap}
\end{align}
The state $\ket{{--'}}$ and its negativity $N(\ket{{--'}})$ are calculated in the first order of the perturbation ${H'_{\textrm{LO}} \rightarrow H'_{\textrm{LO}}+ \beta H_{\textrm{int}}}$. 
The zeroth-order eigenstates and eigenvalues are ${\{\ket{{--}},\ket{{-+}},\ket{{+-}},\ket{{++}}\}}$ and ${\{-2 h'-2 \zeta h',2 \zeta h',2h'\}}$, respectively.
The first-order eigenstate for the state ${\ket{{--}}}$ is then given by
\begin{align}
\ket{{--'}} = \ket{{--}} + \beta n_1 \ket{{-+}} +\beta n_2 \ket{{+-}} +\beta n_3 \ket{{++}}+ O(\beta^2),
\end{align}  
where
\begin{align}
n_1= \frac{\langle {-+} | H_{\textrm{int}} | {--} \rangle}{-2(1-\zeta) h'} ,    \notag \\
n_2=\frac{\langle {+-} | H_{\textrm{int}}  | {--} \rangle}{-2 (\zeta+1) h'}, \notag \\
n_3=\frac{\langle {++} | H_{\textrm{int}}  | {--} \rangle}{-4 h'}.   \label{minus}
\end{align}
Note that the normalization factor of the state $\ket{{--}'}$ is ${1+O(\beta^2)}$.
The matrix representation of ${\rho_0^{T_1}=\bigl(\ket{{--'}}\bra{{--'}}\bigr)^{T_1}}$ is therefore given in the basis of ${\{\ket{{--}},\ket{{-+}},\ket{{+-}},\ket{{++}}\}}$ as follows by ignoring the terms of $O(\beta^2)$:
\begin{align}
\bigl(\ket{{--'}}\bra{{--'}}\bigr)^{T_1} =\begin{pmatrix} 1 &0&0&0  \\ 0&0&0&0 \\ 0 &0 &0&0 \\ 0&0&0&0  \end{pmatrix} + \beta\begin{pmatrix} 0 &n_1^{\ast} &n_2  &0  \\ n_1&0& n_3  &0 \\ n_2^{\ast}&n_3^{\ast} &0&0 \\ 0&0&0&0  \end{pmatrix} . \label{partialstate}
\end{align}
The zeroth-order eigenvalues of $\rho_0^{T_1}$ are ${\{1,0,0,0\}}$. 
The negative eigenvalue emerges when the degeneracy of the first and second zero eigenvalues resolve in the first order of $\beta$.
The third zero eigenvalue remains to be zero. The eigenvalues are then given by ${\{1,\beta |n_3|,-\beta |n_3|,0\}}$ in the first order and hence the negative eigenvalue $-\beta|n_3|$ gives the negativity
\begin{align}
N(\ket{{--'}})&= 2\beta |n_3| \notag \\
&=\beta  \frac{|\langle {++} | H_{\textrm{int}}  | {--} \rangle|}{2h'}   + O(\beta^2), \label{neganega}
\end{align}
The corresponding eigenstate $\ket{\phi_{-}}$ is given by
\begin{align}
\ket{\phi_-} = \frac{1}{\sqrt{2}} \Bigl(\ket{{-+}}-\frac{n_3}{|n_3|}\ket{{+-}} \Bigr)+O(\beta)\ket{{--}}. \label{negastate}
\end{align}
Similarly, we have
\begin{align} 
\rho_{-+'}^{T_1} &= \bigl(\ket{{-+'}} \bra{{-+'}} \bigr)^{T_1}= \ket{{-+}} \bra{{-+}} + O(\beta), \notag \\
\rho_{+-'}^{T_1} &= \bigl(\ket{{+-'}} \bra{{+-'}} \bigr)^{T_1} =\ket{{+-}} \bra{{+-}} + O(\beta), \notag \\
\rho_{++'}^{T_1} &= \bigl(\ket{{++'}} \bra{{++'}} \bigr)^{T_1}  = \ket{{++}} \bra{{++}} + O(\beta), \label{forapp1}
\end{align}
as well as
\begin{align} 
\lambda_{-+'} &= \frac{e^{2\zeta h'-\beta \delta \epsilon_2}}{Z} \notag \\
&\simeq e^{-2(1-\zeta)  h'}(1+ \beta \delta \epsilon_1 )(1- \beta \delta \epsilon_2 )(1-e^{-2(1-|\zeta|)h'}) \notag \\
&=e^{-2(1-\zeta)  h'} +O(\beta^{2-\zeta-|\zeta|}), \notag \\
\lambda_{+-'}&= \frac{e^{-2\zeta h'-\beta \delta \epsilon_3}}{Z} \simeq e^{-2(1+\zeta)  h'} +O(\beta^{2+\zeta-|\zeta|}) , \notag \\
\lambda_{++'} &= \frac{e^{-2 h'-\beta \delta \epsilon_4}}{Z} \simeq e^{-4  h'} +O(\beta^{3-|\zeta|}), \label{forapp2}
\end{align}
where we used Eq.~\eqref{eigenvaZ} for $Z$.
Note that the first term of each of $\lambda_{-+'}$, $\lambda_{+-'}$ and $\lambda_{++'}$ is of order $\beta^{1-\zeta}$, $\beta^{1+\zeta}$ and $\beta^{2}$ or less, respectively, in the range of (\ref{Lemma2}), ${h'> (\log1/\beta)/2}$.
By substituting Eqs.~\eqref{neganega}--\eqref{forapp2} in Eq.~\eqref{negativityap}, we have
\begin{align}
N\biggl(\sum_{i=1}^4 \lambda_i \rho_i \biggr) \simeq& \beta \frac{|\langle {++} | H_{\textrm{int}}  | {--} \rangle|}{2h'}  \notag \\
&- e^{-2(1-\zeta) h'} - e^{-2(1+\zeta) h'} +O(\beta^{2(1-|\zeta|)}) \label{negativityappro}
\end{align}
for ${\zeta\neq\pm1}$.

Because the matrix element ${|\langle {++} | H_{\textrm{int}}  | {--} \rangle |}$ is independent of $h'$ and $\zeta$, we can solve the maximization problem of Eq.~(\ref{negativityappro}) as follows. First, to maximize the negative terms in Eq.~(\ref{negativityappro}), we must put ${\zeta =0}$.
Then, by differentiating Eq.~(\ref{negativityappro}) with $h'$, we have the optimizing parameter $h'_{\textrm{op}}$ as a solution of
\begin{align}
e^{2h'_{\textrm{op}}} \simeq \frac{8h'^2_{\textrm{op}}}{\beta|\langle {++} | H_{\textrm{int}}  | {--} \rangle |}.  \label{happro2} 
\end{align}
The optimized negativity is then given by 
\begin{align}
N_{\textrm{op}} (\rho) &\simeq\beta \frac{|\langle {++} | H_{\textrm{int}}  | {--} \rangle |}{2h'_{\textrm{op}}} - 2e^{-2 h'_{\textrm{op}} } \notag \\
 &\simeq \beta \frac{|\langle {++} | H_{\textrm{int}}  | {--} \rangle |}{2h'_{\textrm{op}} }  \Bigl(1- \frac{1}{2h'_{\textrm{op}} } \Bigr)    ,                \label{Nappro2}
\end{align}
where we used Eq.~(\ref{happro2}) upon moving from the first line to the second line. 
This is the result for ${\zeta \neq \pm1}$.
From Lemma~3, we see that the optimized negativity \eqref{Nappro2} in the case of ${\zeta = 0}$ is larger in the limit ${\beta \rightarrow 0}$ than the one \eqref{degenerate} in the cases of ${\zeta = \pm1}$.

The other optimizing parameters to be fixed are ${\{\theta_1,\phi_1,\theta_2,\phi_2\}}$.
Let us see how these parameters affect the value of (\ref{Nappro2}).
These parameters affect the matrix element ${|\langle {++} | H_{\textrm{int}}  | {--} \rangle |}$ and hence the value of (\ref{Nappro2}) directly as well as indirectly through $h'_{\textrm{op}}$ given by Eq.~(\ref{happro2}).
We can write down the solution of Eq.~(\ref{happro2}) in terms of Lambert's W function~\cite{Wfunction}, which is defined as a solution of 
\begin{align}
x = W(x) e^{W(x)},
\end{align}
because we can cast Eq.~(\ref{happro2}) into the form
\begin{align}
 (-h'_{\textrm{op}})e^{-h'_{\textrm{op}}}  =- \sqrt{ \frac{\beta|\langle {++} | H_{\textrm{int}}  | {--} \rangle |}{8}}.  \label{happro22} 
\end{align}
The appropriate solution of Eq.~(\ref{happro2}) is given by
\begin{align}
h'_{\textrm{op}} \simeq  - W_{-1} \biggl( -\sqrt{\frac{\beta |\langle {++} | H_{\textrm{int}}  | {--} \rangle |}{8}} \biggr), \label{wfunchop}
\end{align}
where $W_{-1}(x)$ is the branch of ${W(x)}$ satisfying ${W_{-1}(x)\le -1}$ in the domain ${-1/e<x<0}$~\cite{Wfunction}.
The function ${-W_{-1} (-x)}$ is a monotonically decreasing function of $x$ in the domain ${0<x<1/e}$.
Therefore, maximizing the element ${|\langle {++} | H_{\textrm{int}}  | {--} \rangle |}$ with respect to the parameters ${\{\theta_1,\phi_1,\theta_2,\phi_2\}}$ brings $h'_{\textrm{op}}$ to its minimum within the range ${h'_{\textrm{op}} >1}$.
Since the factor
\begin{align}
\frac{1}{2h'_{\textrm{op}} }  \Bigl(1- \frac{1}{2h'_{\textrm{op}} } \Bigr)               \label{factor1}     
\end{align}
in Eq.~(\ref{Nappro2}) is a decreasing function of $h'_{\textrm{op}}$ for ${h'_{\textrm{op}}>1}$, minimizing $h'_{\textrm{op}}$ within the range ${h'_{\textrm{op}}>1}$ brings the factor (\ref{factor1}) to its maximum.
To summarize, the element ${|\langle {++} | H_{\textrm{int}}  | {--} \rangle |}$ increases the value of (\ref{Nappro2}) not only directly but also through $h'_{\textrm{op}}$ indirectly.

The next task is then to find the parameters ${\{\theta_1,\phi_1,\theta_2,\phi_2\}}$ that maximize the matrix element ${|\langle {++} | H_{\textrm{int}}  | {--} \rangle |}$ in {Eq.~(\ref{Nappro2})}.
The eigenstates of the one-qubit part ${\sum_{i = x,y,z}  h^i \sigma^i}$ of the local Hamiltonian $H_{\textrm{LO}}$ are given by
\begin{align}
\ket{+}&=\cos \frac{\theta}{2}\ket{0} + e^{i \phi} \sin \frac{\theta}{2}\ket{1} ,\notag \\
\ket{-}&=-\sin \frac{\theta}{2}\ket{0} + e^{i \phi} \cos \frac{\theta}{2}\ket{1} ,
\end{align}
where we define $\ket{0}$ and $\ket{1}$ as the eigenstates of $\sigma^z$ and represent $\{h^i\}_{i=x,y,z}$ as ${\{h \sin \theta \cos \phi ,h \sin \theta \sin \phi,h\cos \theta\}}$ in the polar coordinate.
We can thereby express the eigenstates $\ket{{++}}$ and $\ket{{--}}$ of $H_{\textrm{LO}}$ in the forms 
\begin{align}
\ket{{++}}&=\cos \frac{\theta_1}{2} \cos \frac{\theta_2}{2}\ket{00} + \cos \frac{\theta_1}{2}\sin \frac{\theta_2}{2}e^{i \phi_2}\ket{01} \notag \\
                 &+\sin \frac{\theta_1}{2}\cos \frac{\theta_2}{2}e^{i \phi_1} \ket{10}+\sin \frac{\theta_1}{2}\sin \frac{\theta_2}{2}e^{i (\phi_1+\phi_2)} \ket{11} ,\notag \\
\ket{{--}}&=\sin \frac{\theta_1}{2} \sin \frac{\theta_2}{2}\ket{00} - \sin \frac{\theta_1}{2}\cos \frac{\theta_2}{2}e^{i \phi_2}\ket{01} \notag \\
               &-\cos \frac{\theta_1}{2}\sin \frac{\theta_2}{2}e^{i \phi_1} \ket{10}+\cos \frac{\theta_1}{2}\cos \frac{\theta_2}{2}e^{i (\phi_1+\phi_2)} \ket{11} .
\end{align}
We therefore have the matrix element ${\langle {++} | H_{\textrm{int}} | {--} \rangle}$ in the following form:
\begin{align}
&\langle {++} | H_{\textrm{int}} | {--} \rangle \notag \\
=&J_z \sin \theta_1 \sin \theta_2 + (J_x-J_y)\Bigl[\sin^2 \frac{\theta_1}{2}\sin^2 \frac{\theta_2}{2}e^{-i (\phi_1+\phi_2)} \notag \\
+& \cos^2 \frac{\theta_1}{2}\cos^2 \frac{\theta_2}{2}e^{i (\phi_1+\phi_2)}\Bigr]  \notag \\
 -& (J_x+J_y)\Bigl[\cos^2 \frac{\theta_1}{2}\sin^2 \frac{\theta_2}{2}e^{i (\phi_1-\phi_2)} \notag \\
  +& \sin^2 \frac{\theta_1}{2}\cos^2 \frac{\theta_2}{2}e^{i (-\phi_1+\phi_2)}\Bigr] .
\end{align}

In the cases of ${\{J_{x},J_{y}\}\ge J_{z}\ge0}$ and ${0\ge J_{z} \ge \{J_{x},J_{y}\}}$, the upper bound of ${|\langle {++} | H_{\textrm{int}}  | {--} \rangle|}$ is given by
\begin{align}
|\langle {++} | H_{\textrm{int}}  | {--} \rangle | \le |J_x+J_y|. \label{inequalitypm}
\end{align}
We prove this inequality in the cases of ${J_{x}\ge J_{y}\ge J_{z}\ge0}$; we can prove the other cases in the same way.
First, ${|\langle {++} | H_{\textrm{int}}  | {--} \rangle |}$ satisfies the following inequality:
\begin{align}
&|\langle {++} | H_{\textrm{int}} | {--} \rangle|  \notag \\
\le& |J_z| \sin \theta_1 \sin \theta_2  \notag \\
&+ |J_x-J_y|\Bigl(\sin^2 \frac{\theta_1}{2}\sin^2 \frac{\theta_2}{2}
+ \cos^2 \frac{\theta_1}{2}\cos^2 \frac{\theta_2}{2} \Bigr)  \notag \\
&+ |J_x+J_y| \Bigl(\cos^2 \frac{\theta_1}{2}\sin^2 \frac{\theta_2}{2} 
  + \sin^2 \frac{\theta_1}{2}\cos^2 \frac{\theta_2}{2}\Bigr) \notag  \\
=& |J_z| \sin \theta_1 \sin \theta_2  + |J_x-J_y|\frac{1+ \cos \theta_1 \cos \theta_2}{2} \notag \\
&\ \ \ \ \ \ \ \ \ \ \ \ \ \ \ \ \ \ \ + |J_x+J_y| \frac{1- \cos \theta_1 \cos \theta_2}{2} . \label{plusminus}
\end{align}
By utilizing the fact that ${J_{x}\ge J_{y}\ge J_{z}\ge0}$, the inequality~\eqref{plusminus} reduces to
\begin{align}
|\langle {++} | H_{\textrm{int}} | {--} \rangle|  \le& J_x - J_y \cos \theta_1 \cos \theta_2  + J_z\sin \theta_1 \sin \theta_2  \notag \\
\le& |J_x+J_y|.
\end{align}
The inequality~\eqref{inequalitypm} becomes an equality when we choose ${\{\theta_1,\theta_2, \phi_1,\phi_2\}}$ as ${\{0,\pi,0,0\}}$ for example, or 
 in the Cartesian coordinate $\{h_1^x,h_1^y,h_1^z,h_2^x,h_2^y,h_2^z\}= \{0,0,h,0,0,-h\}$. Then, the optimizing local parameters are given in the form of $\{h_1^x,h_1^y,h_1^z,h_2^x,h_2^y,h_2^z\} = \{0,0, h_{\textrm{op}},0,0,-h_{\textrm{op}}\}$. Moreover, Eqs.~(\ref{happro}) and (\ref{Nappro}) can be given by substituting $|\langle {++} | H_{\textrm{int}}  | {--} \rangle|$ with $|J_x+J_y|$ in Eqs.~(\ref{happro2}) and (\ref{Nappro2}).
 
Finally, the leading order of Lambert's W function ${-W_{-1} (-x)}$ is ${\log x}$~\cite{Wfunction}. Therefore, the leading order of Eq.~(\ref{wfunchop}) gives Eq.~(\ref{hopappro}),
which then results in Eqs.~(\ref{theorem1}) and (\ref{theorem1N}).
This completes the proof of Theorem~1.

\subsection{Optimization at arbitrary temperatures}
It is difficult to generalize Theorem~1 to arbitrary temperatures. However, we can present the following Theorem~2. Let us now parametrize the local fields as follows:
\begin{align}
&\{h_1^x,h_1^y,h_1^z,h_2^x,h_2^y,h_2^z\} \notag \\
=& \{h_1^x,h_1^y,h^z(1+\xi),h_2^x,h_2^y,-h^z(1-\xi)\},
\end{align}
or
\begin{align}
h^z\equiv \frac{h_1^z-h_2^z}{2},  \quad \xi \equiv \frac{h_1^z+h_2^z}{h_1^z-h_2^z}. 
\end{align}

\textit{Theorem~2}. When we express the negativity as a function of the local parameters ${\{h_1^x,h_1^y,h^z(1+\xi),h_2^x,h_2^y,-h^z(1-\xi)\}}$, the following equation holds at arbitrary temperatures:
\begin{align}
&\frac{\partial N}{ \partial h_1^x} =\frac{\partial N}{ \partial h_2^x}= \frac{\partial N}{ \partial h_1^y} =\frac{\partial N}{ \partial h_2^y} =\frac{\partial N}{ \partial \xi}  =0 \notag \\  
&\textrm{at}\ \{h_1^x,h_1^y,h_2^x,h_2^y,\xi,h^z\}=\{0,0,0,0,0,h\}. \label{theorem2}
\end{align}
This theorem means that the form of the optimizing local parameters in the high-temperature limit, $\{h_1^x,h_1^y,h_1^z,h_2^x,h_2^y,h_2^z\}=\{0,0,h_{\textrm{op}},0,0,-h_{\textrm{op}}\}$, also gives an extremal value of the negativity at arbitrary temperatures.

\textit{Proof}.
To prove this theorem, we firstly calculate the perturbation of the negativity due to an infinitesimal variation of the local parameters at arbitrary temperatures. If it always vanishes, Eq.~(\ref{theorem2}) is proved.
We first derive the perturbation of the density matrix due to an infinitesimal variation of the local parameters, from ${\{0,0,h,0,0,-h\}}$ to $\{\delta h_1^x,\delta h_1^y,h(1+\delta \xi),\delta h_2^x,\delta h_2^y,-h(1- \delta \xi)\}$. This means the perturbation of the form
 \begin{align} 
H_{\textrm{tot}} &= H_{\textrm{tot}}^{\textrm{op}} + \delta \mathcal{H}_{\textrm{LO}},  
\end{align}
where
\begin{align}
H_{\textrm{tot}}^{\textrm{op}}&\equiv \sum_{i= x,y,z} J_{i} \sigma_1^i\otimes \sigma_2^i + h (\sigma_1^z\otimes I -I \otimes \sigma_2^z)  \label{Htotop}
\end{align}
is the total Hamiltonian with the local parameters ${\{0,0,h,0,0,-h\}}$ and
\begin{align}
\delta \mathcal{H}_{\textrm{LO}}  &\equiv \sum_{i = x,y} ( \delta h_1^i \sigma_1^i\otimes I +\delta h_2^i I \otimes \sigma_2^i) \notag \\
&\ \ \ \ \ \ \ \ \ \ +h \delta \xi  (\sigma_1^z\otimes I +I \otimes \sigma_2^z)   \label{setudouH}
\end{align} 
is the infinitesimal variation of the local Hamiltonian.
{Equation~(\ref{perturbation})} gives the perturbation of the density matrix ${\delta \rho}$ as
\begin{align}
\delta \rho &=\frac{ e^{ -\beta (H_{\textrm{tot}}^{\textrm{op}}+\delta \mathcal{H}_{\textrm{LO}} ) }}{Z+\delta Z} -\frac{ e^{-\beta H_{\textrm{tot}}^{\textrm{op}} }}{Z}  \notag  \\
                      &=- \frac{\delta Z}{Z} \rho_{\textrm{op}} - \frac{\beta}{Z} \int_0^1 e^{- \beta (1-x) H_{\textrm{tot}}^{\textrm{op}} } \delta \mathcal{H}_{\textrm{LO}} e^{- \beta x H_{\textrm{tot}}^{\textrm{op}} } dx ,
\end{align}
where ${\rho_{\textrm{op}} =e^{-\beta H_{\textrm{tot}}^{\textrm{op}} }/Z}$ and 
\begin{align}
\delta Z &=  \textrm{tr} \biggl(  -\beta \int_0^1 e^{- \beta (1-x) H_{\textrm{tot}}^{\textrm{op}} } \delta \mathcal{H}_{\textrm{LO}} e^{- \beta x H_{\textrm{tot}}^{\textrm{op}} } dx \biggr ).
\end{align}
Then, the perturbation of the negativity in Eq.~(\ref{negaperturbation}), ${\delta N =- 2\langle \phi_-| \delta \rho^{T_1} | \phi_- \rangle}$, is given as
\begin{align}
\delta N  &=- \frac{\delta Z}{Z} N(\rho_{\textrm{op}})   \notag \\
+\frac{2\beta}{Z}&  \int_0^1 \textrm{tr} \biggl [\ket{ \phi_- }  \bra{ \phi_- }\Bigl (e^{- \beta (1-x) H_{\textrm{tot}}^{\textrm{op}} } \delta \mathcal{H}_{\textrm{LO}} e^{- \beta x H_{\textrm{tot}}^{\textrm{op}} } \Bigr)^{T_1} \biggr] dx \notag \\
&=\frac{\beta N(\rho_{\textrm{op}}) }{Z}  \textrm{tr} \biggl( \int_0^1 e^{- \beta (1-x) H_{\textrm{tot}}^{\textrm{op}} } \delta \mathcal{H}_{\textrm{LO}} e^{- \beta x H_{\textrm{tot}}^{\textrm{op}} } dx \biggr ) \notag \\
+\frac{2\beta}{Z}& \int_0^1\textrm{tr}\Bigl [  \bigl(\ket{ \phi_- } \bra{ \phi_- }\bigr)^{T_1}  e^{- \beta (1-x) H_{\textrm{tot}}^{\textrm{op}} }  \delta \mathcal{H}_{\textrm{LO}}  e^{- \beta x H_{\textrm{tot}}^{\textrm{op}} } \Bigr] dx  \notag \\
&= \int_0^1\textrm{tr}\Bigl [ e^{- \beta x H_{\textrm{tot}}^{\textrm{op}} }  \hat{n}  e^{- \beta (1-x) H_{\textrm{tot}}^{\textrm{op}} }  \delta \mathcal{H}_{\textrm{LO}} \Bigr ] dx      \label{pertrace1}
\end{align}
where 
\begin{align}
\hat{n} \equiv  \frac{\beta}{Z}\Bigl[  N(\rho_{\textrm{op}}) \bigl( I\otimes I \bigr)  +2\bigl(\ket{ \phi_- } \bra{ \phi_- }\bigr)^{T_1}\Bigr]  
\end{align}
 and we used the identity ${N(\rho_{\textrm{op}}) =- 2 \langle \phi_-| \rho_{\textrm{op}}^{T_1} | \phi_- \rangle}$ as well as ${\textrm{tr} (A^{T_1} B) =\textrm{tr} (A B^{T_1})}$. We will prove that the integrand of Eq.~\eqref{pertrace1},
\begin{align}
\textrm{tr}&\Bigl[ e^{- \beta x H_{\textrm{tot}}^{\textrm{op}} } \hat{n} e^{- \beta (1-x) H_{\textrm{tot}}^{\textrm{op}} }
 \delta \mathcal{H}_{\textrm{LO}}  \Bigr]   ,\label{pertrace2}
\end{align}
always vanishes for $\{h_1^x,h_1^y,h_2^x,h_2^y,\xi,h_z\}=\{0,0,0,0,0,h \}$.

We prove in Appendix~C that the operator 
\begin{align}
e^{- \beta x H_{\textrm{tot}}^{\textrm{op}} } \hat{n} e^{- \beta (1-x) H_{\textrm{tot}}^{\textrm{op}} }  \label{setudoukou}
\end{align}
has the same symmetry as the Hamiltonian $H_{\textrm{tot}}^{\textrm{op}}$ in Eq.~(\ref{Htotop}), and thereby must be expanded in terms of the Pauli matrices in the form
\begin{align}
\frac{1}{4} \biggl(q_{00} I \otimes I + q_{z0}(\sigma_1^z \otimes I -   I \otimes \sigma_2^z)
                                  + \sum_{i=x,y,z} q_{ii} \sigma_1^i\otimes \sigma_2^i  \biggr), \label{taisyou}
\end{align}
where $q_{00}$, $q_{z0}$ and $q_{ii}$ are appropriate coefficients.  
Therefore, we can calculate Eq.~(\ref{pertrace2}) to have the following equation:
\begin{align}
&\textrm{tr}\Bigl[ e^{- \beta x H_{\textrm{tot}}^{\textrm{op}} } \hat{n}  e^{- \beta (1-x) H_{\textrm{tot}}^{\textrm{op}} }  \delta \mathcal{H}_{\textrm{LO}}  \Bigr]  \notag \\
=&\textrm{tr}\Biggl\{ \frac{1}{4} \biggl[q_{00} I \otimes I + q_{z0}(\sigma_1^z \otimes I -   I \otimes \sigma_2^z)  \notag \\
                                  &+ \sum_{i=x,y,z} q_{ii} \sigma_1^i\otimes \sigma_2^i  \biggr] \times \notag \\
                                \biggl[\sum_{i = x,y}& ( \delta h_1^i \sigma_1^i\otimes I +\delta h_2^i I \otimes \sigma_2^i) +h \delta \xi  (\sigma_1^z\otimes I +I \otimes \sigma_2^z)\biggr] \Biggl \}.       \label{appeC}                        
\end{align}
A straightforward algebra, such as ${\textrm{tr}(\sigma_1^z \otimes \sigma_2^x)=0}$, yields that Eq.~(\ref{appeC}) vanishes.    
This means that the perturbation of the negativity due to the infinitesimal variation of the local parameters $\{\delta h_x^1,\delta h_y^1,\delta h_x^2,\delta h_y^2, \delta \xi\}$ always vanishes at $\{h_1^x,h_1^y,h_2^x,h_2^y,\xi,h_z\}=\{0,0,0,0,0,h\}$. This completes the proof of Theorem~2.

To extend Theorem~1 to arbitrary temperatures, we assume the following hypothesis:\\
\textit{Hypothesis~1. The local parameters of the form $\{h_1^x,h_1^y,h_1^z,h_2^x,h_2^y,h_2^z\}=\{0,0,h_{\textrm{op}},0,0,-h_{\textrm{op}}\}$ give not only an extremal value but also the maximum value of entanglement at arbitrary temperatures. }\\
We numerically tested this hypothesis using determinant-based entanglement measure ${\pi (\rho)}$~\cite{Augusiak}, which is given as
\begin{align}
\pi (\rho) \equiv \begin{cases}
                            0,  & \textrm{for}\ \rho^{T_1} \ge 0 , \\
                            2(|\det \rho^{T_1} |)^{1/4},  & \textrm{for}\ \rho^{T_1} < 0 .                                                        
                              \end{cases} 
\end{align}
Though this entanglement measure is not a full entanglement monotone, it provides tight lower and upper bounds for other entanglement measures including the negativity and the concurrence. In addition, ${\det \rho^{T_1}}$ is expressed in the form of a polynomial and is much easier to maximize numerically than the concurrence and the negativity. Utilizing this measure, we tested Hypothesis~1 by numerical optimization for various kinds of interaction at various temperatures and found it always satisfied. In the following, we will assume Hypothesis~1 and conclude that $\{h_1^x,h_1^y,h_1^z,h_2^x,h_2^y,h_2^z\}=\{0,0,h_{\textrm{op}},0,0,-h_{\textrm{op}}\}$ is the globally optimizing solution at any temperatures.

For the local parameters $\{h_1^x,h_1^y,h_1^z,h_2^x,h_2^y,h_2^z\}=\{0,0,h,0,0,-h\}$, the density matrix $Z\rho^{T_1}$ is given at arbitrary temperatures in the basis of the eigenstates of ${\sigma_1^z\otimes \sigma_2^z}$, ${\{\ket{{00}},\ket{{01}},\ket{{10}},\ket{{11}}\}}$, as
\begin{align}
Z\rho^{T_1} = \begin{pmatrix} a_1 &0&0& a_2   \\ 0 &b_1 -b_2 &b_3&0 \\ 0&b_3&b_1 +b_2&0 \\ a_2&0&0&a_1  \end{pmatrix} , \label{rhotmat}
\end{align}
where
\begin{align}
a_1 &= e^{-\beta J_z} \cosh \beta J_1 , \ \ 
a_2 = -\frac{e^{\beta J_z}  (J_x + J_y) \sinh \beta J_2 }{J_2}, \notag \\
b_1 &=  e^{\beta J_z} \cosh \beta J_2 , \ \ 
b_2 = \frac{2h e^{\beta J_z} \sinh \beta J_2 }{J_2}, \notag \\
b_3 &= - e^{-\beta J_z} \sinh \beta J_1, \notag \\
J_1 &\equiv |J_x-J_y| ,  \ \
J_2 \equiv\sqrt{4h^2 + (J_x+J_y)^2} .
\end{align}
Its eigenvalues are 
\begin{align}
\Bigl \{a_1-|a_2|,a_1+|a_2|,b_1+\sqrt{b_2^2+b_3^2},b_1-\sqrt{b_2^2+b_3^2} \Bigr \}. \label{rhoeigenvalues}
\end{align}
In Appendix~D, we will prove that only ${a_1-|a_2|}$ can have a negative value for ${\{J_{x},J_{y}\}\ge J_{z}\ge0}$ and ${0\ge J_{z} \ge \{J_{x},J_{y}\}}$.
Therefore, the optimized negativity is given by
\begin{align}
N(J_x,J_y,J_z,h,\beta)=\textrm{max}(\tilde{N},0) , \label{optnega}
\end{align}
where
\begin{align}
\tilde{N} &= -2\frac{a_1- |a_2|}{Z}\notag \\
&= -\frac{ e^{-\beta J_z} \cosh \beta J_1 -  \bigl( e^{\beta J_z} |J_x+J_y| \sinh \beta J_2 \bigr) /J_2 }{e^{-\beta J_z}  \cosh \beta J_1 + e^{\beta J_z} \cosh \beta J_2 }  , \notag \\
Z&=2e^{-\beta J_z} \cosh \beta J_1 + 2e^{\beta J_z} \cosh \beta J_2 . \label{optimizenega}
\end{align}
We find from this expression that we can always make the negativity positive by choosing an appropriate value of $h$.

The remaining task is to find the value of the optimizing field $h_{\textrm{op}}$ at each temperature. We will do it analytically in the low-temperature limit ${\beta \rightarrow \infty}$ in Sec.~III.C as well as do it numerically rigorously for a wide range of the temperature in Sec.~IV.

\subsection{Optimization in the low-temperature limit}
We now discuss the optimization problem in the low-temperature limit. 

\textit{Theorem~3}.
 In the low-temperature limit ${\beta\rightarrow \infty}$, the optimized entanglement approaches to $1$. The optimizing parameter $h_{\textrm{op}}$ approaches to $0$ when we choose the optimizing parameters as ${\{0,0,h_{\textrm{op}},0,0,-h_{\textrm{op}}\}}$.
 
\textit{Proof}. 
We need to consider the three cases, namely the cases where the ground state of $H_{\textrm{int}}$ is non-degenerate, doubly degenerate and triply degenerate.
The eigenvalues $\{ \epsilon_i \}_{i=1}^4$ and the corresponding eigenstates $\{ \ket{\psi_i} \}_{i=1}^4$ of $H_{\textrm{int}}$ are given by the following:
\begin{align}
H_{\textrm{int}} &= \sum_{i= x,y,z} J_{i} \sigma_1^i \otimes \sigma_2^i, \notag \\
\epsilon_1 &= -J_x - J_y-J_z,\ \  \ket{\psi_1} = \frac{1}{\sqrt{2}}\Bigl ( \ket{01} -\ket{10}  \Bigr) ,\notag \\
\epsilon_2 &= J_x + J_y-J_z, \ \  \ket{\psi_2} = \frac{1}{\sqrt{2}}\Bigl ( \ket{01}+\ket{10}  \Bigr),  \notag \\
\epsilon_3 &= J_x - J_y+J_z  ,  \ \  \ket{\psi_3} = \frac{1}{\sqrt{2}}\Bigl ( \ket{00}+\ket{11}  \Bigr), \notag \\
\epsilon_4 &= -J_x + J_y+J_z  ,  \ \  \ket{\psi_4} = \frac{1}{\sqrt{2}}\Bigl ( \ket{00}-\ket{11}  \Bigr). 
\end{align}
As has been described in Sec.~II, we consider only the cases of ${\{J_{x},J_{y}\}\ge J_{z}\ge0}$ and ${0\ge J_{z} \ge \{J_{x},J_{y}\}}$. 

In each case of ${\{J_{x},J_{y}\}\ge J_{z}\ge0}$ or ${0 \ge J_{z} > \{J_{x},J_{y}\}}$, the ground state of $H_{\textrm{int}}$ is non-degenerate, and $\epsilon_1$ or $\epsilon_2$ is the ground-state eigenvalue, respectively. 
In these cases, the ground state is a Bell state and it is clear that its entanglement is maximum. In other words, there is no need to optimize it further and ${H_{\textrm{LO}}^{\textrm{op}}=0}$. We will see in Sec.~IV that, in this non-degenerate case, there is indeed a finite range of the temperature where the negativity is maximized for ${H_{\textrm{LO}}^{\textrm{op}}=0}$.

In each case of ${0 \ge J_{z} = J_{x} > J_{y}}$ and ${0 \ge J_{z} = J_{y} > J_{x}}$, the ground state of $H_{\textrm{int}}$ is doubly degenerate and ${\epsilon_2=\epsilon_4}$ or ${\epsilon_2=\epsilon_3}$ is the ground-state eigenvalue, respectively. 
In the case ${0 \ge J_{z} = J_{x} = J_{y}}$, the ground state of $H_{\textrm{int}}$ is triply degenerate and ${\epsilon_2=\epsilon_3=\epsilon_4}$ is the ground-state eigenvalue. 
In these degenerate cases, the ground states are mixed states and their entanglement always vanish. However, we can resolve the degeneracy of the ground states by an infinitesimal local Hamiltonian.

We hence employ Hypothesis~1 and put $\{h_1^x,h_1^y,h_1^z,h_2^x,h_2^y,h_2^z\}=\{0,0,h_{\textrm{op}},0,0,-h_{\textrm{op}}\}$. We then calculate the asymptotic behavior of the optimized entanglement in the low-temperature limit ${\beta\rightarrow \infty}$.
Below we will derive
\begin{align} 
h_{\textrm{op}} &\simeq \sqrt{\frac{\tilde{J} }{2\beta} \log2 \beta\tilde{J}  } \quad \mbox{as} \quad  \beta\rightarrow \infty, \notag \\
N_{\textrm{op}} &\simeq 1- \frac{1+ \log 2 \beta \tilde{J} }{\beta \tilde{J} }  \quad \mbox{as} \quad  \beta\rightarrow \infty. \label{doubledegenerate}
\end{align}
in the doubly degenerate cases, where we defined ${\tilde{J} \equiv |J_x+J_y|}$,
and
\begin{align} 
h_{\textrm{op}} &\simeq \sqrt{\frac{\tilde{J} }{2\beta} \log 4 \beta \tilde{J}  } \quad \mbox{as} \quad  \beta\rightarrow \infty \notag \\
N_{\textrm{op}} &\simeq 1- \frac{1+ \log 4 \beta \tilde{J} }{\beta \tilde{J} } \quad \mbox{as} \quad  \beta\rightarrow \infty. \label{tripledegenerate}
\end{align}
in the triply degenerate case.
In both cases the optimizing parameter $h_{\textrm{op}}$ is infinitesimal and the optimized negativity $N_{\textrm{op}}$ approaches to $1$ in the low-temperature limit ${\beta\rightarrow \infty}$, although the forms of $h_{\textrm{op}}$ and $N_{\textrm{op}}$ are slightly different in the two cases. We will see in Sec.~IV that, in these degenerate cases, there is indeed \textit{no} finite range of the temperature where the negativity is maximized without local fields. In other words, we need a non-zero value of $h_{\textrm{op}}$ at any non-zero temperatures.

Now we derive Eqs.~(\ref{doubledegenerate}) and (\ref{tripledegenerate}). 
We start from Eq.~(\ref{optimizenega}) under Hypothesis~1. In the doubly degenerate cases ${0 \ge J_{z} = J_{x} > J_{y}}$ and ${0 \ge J_{z} = J_{y} > J_{x}}$, we can approximate Eq.~(\ref{optimizenega}) as
\begin{align}
\tilde{N} &\simeq -\frac{ e^{-\beta (J_z-J_1)}  -  \bigl( e^{\beta (J_z+J_2)  } |J_x+J_y|\bigr) /J_2 }{e^{-\beta (J_z -J_1)} + e^{\beta (J_z +J_2)} }  \notag \\
&=\frac{ -1 +  \bigl( e^{\beta (2J_z+J_2-J_1)  } |J_x+J_y|\bigr) /J_2 }{1 + e^{\beta (2J_z+J_2-J_1) } }  \label{doubledeg}
\end{align}
in the low-temperature limit ${\beta \rightarrow \infty}$, where we used the facts that ${2\cosh \beta J_1 \simeq e^{\beta J_1}}$, ${2\sinh \beta J_2 \simeq e^{\beta J_2}}$ and ${2\cosh \beta J_2 \simeq e^{\beta J_2}}$.
Moreover, in these doubly degenerate cases, ${2J_z+J_2-J_1}$ is either ${2J_z+J_2-J_x+J_y}$ or ${2J_z+J_2+J_x-J_y}$, which are summarized to ${J_2-|J_x + J_y|}$. Then Eq.~(\ref{doubledeg}) reduces to
\begin{align}
\tilde{N} \simeq \frac{ -1 + e^{\beta X  } \tilde{J} /(X + \tilde{J}) }{1 + e^{\beta X } } , \label{Ntilde}
\end{align}
where
\begin{align}
\tilde{J} &=  |J_x+J_y|, \notag \\ 
X&\equiv J_2 - \tilde{J} = \sqrt{4h^2 + (J_x+J_y)^2}- |J_x+J_y| . \label{JandX}
\end{align}

We first prove that ${X \rightarrow 0}$ and ${\beta X \rightarrow \infty}$ is a necessary and sufficient condition for ${\tilde{N} \rightarrow 1}$ in the low-temperature limit ${\beta \rightarrow \infty}$.
In order to prove this, we calculate the value of ${1-\tilde{N}}$ as follows:
\begin{align}
1- \tilde{N} &=1-  \frac{ -1 + e^{\beta X  } \tilde{J} /(X + \tilde{J}) }{1 + e^{\beta X } }  \notag \\
&=\frac{ 2e^{-\beta X  } +  X /(X + \tilde{J}) }{1 + e^{-\beta X } }  , 
\end{align}
Because ${X\ge 0}$ and ${0< e^{-\beta X} \le 1}$,
we have ${X /(X + \tilde{J}) \ge 0}$ and ${1< 1 + e^{-\beta X } \le 2}$. Therefore, the necessary and sufficient condition for ${1- \tilde{N} \rightarrow 0}$ in the low-temperature limit is
\begin{align}
\beta X \rightarrow \infty  \quad  \mbox{and}\quad X \rightarrow 0 \quad \mbox{as} \quad  \beta\rightarrow \infty. \label{N1condition}
\end{align}
In such cases, the negativity can be maximized to $1$ in the low temperature limit ${\beta  \rightarrow \infty}$. 

Let us now calculate the optimizing parameter ${X_{\textrm{op}}}$. From the extremal condition for Eq.~\eqref{Ntilde}, 
\begin{align}
\frac{d\tilde{N}}{dX} = \frac{e^{\beta X} (\beta X^2 +3  \beta \tilde{J}  X  +2 \beta \tilde{J}^2   -\tilde{J} -e^{\beta X})}{(1+e^{\beta X})^2(\tilde{J} +X)^2}= 0, \label{derivative}
\end{align}
we obtain
\begin{align}
\beta X_{\textrm{op}}= \log \Bigl(\frac{\beta X_{\textrm{op}}^2 }{\tilde{J}}+ 3  \beta X_{\textrm{op}}+  2 \beta \tilde{J} -1    \Bigr) . \label{dxdN}
\end{align}
Because of the condition \eqref{N1condition}, Eq.~(\ref{dxdN}) reduces to
\begin{align}
\beta X_{\textrm{op}}&= \log 2 \beta \tilde{J} + \log\Bigl( 1+  \frac{ X_{\textrm{op}}^2 }{2\tilde{J}^2}+\frac{3 X_{\textrm{op}}   }{2\tilde{J}}  - \frac{1}{2\beta \tilde{J}}   \Bigr) \notag \\
           &\simeq  \log 2\beta  \tilde{J}  \label{Xoptap}
\end{align}
in the limit ${\beta  \rightarrow \infty}$. We thus have
\begin{align}
X_{\textrm{op}}  \simeq \frac{\log 2\beta  \tilde{J}}{\beta} , \label{Xopt}
\end{align}
which indeed satisfies \eqref{N1condition}.
The optimizing parameter $h_{\textrm{op}}$ is thereby obtained in the form 
\begin{align}
h_{\textrm{op}} &=\frac{1}{2}\sqrt{X_{\textrm{op}}^2 + 2 \tilde{J} X_{\textrm{op}}} \notag \\
      &\simeq \sqrt{ \frac{\tilde{J} X_{\textrm{op}}}{2}} \notag \\
      &\simeq \sqrt{\frac{\tilde{J} }{2\beta} \log2 \beta\tilde{J}  },
\end{align}
where we utilized Eq.~(\ref{JandX}) to derive the first equality.
Moreover, the optimized negativity is given by
\begin{align}
N_{\textrm{op}} &\simeq \frac{ -e^{-\beta X _{\textrm{op}}}  + 1 /(X_{\textrm{op}}/ \tilde{J} +1) }{ e^{-\beta X _{\textrm{op}}} +1 } \notag \\
&\simeq (1-e^{-\beta X _{\textrm{op}}} )\biggl( -e^{-\beta X _{\textrm{op}}}  + 1- \frac{X_{\textrm{op}}}{ \tilde{J} } \biggr)  \notag \\
&\simeq 1- \frac{X_{\textrm{op}}}{ \tilde{J} } - 2e^{-\beta X _{\textrm{op}}} \notag \\
&\simeq 1- \frac{1+ \log 2 \beta \tilde{J} }{\beta \tilde{J} },
\end{align}
where we used  Eq.~\eqref{N1condition} upon moving from the first line to the second line.
Thus Eq.~(\ref{doubledegenerate}) is proved.

In the triply degenerate case ${0 \ge J_{z} = J_{x} =J_{y}}$, we have ${J_1=0}$, and thereby we can approximate Eq.~(\ref{optimizenega}) as
\begin{align}
\tilde{N} &\simeq -\frac{ 2 e^{-\beta J_z}  -  \bigl( e^{\beta (J_z+J_2)  } |J_x+J_y|\bigr) /J_2 }{2 e^{-\beta J_z } + e^{\beta (J_z +J_2)} }  \notag \\
&=\frac{ 2 +  \bigl( e^{\beta (2J_z+J_2)  } |J_x+J_y|\bigr) /J_2 }{2 + e^{\beta (2J_z+J_2) } }\label{tripledeg}
\end{align}
in the low-temperature limit ${\beta \rightarrow \infty}$, where we used the facts that ${\cosh \beta J_1 \simeq 1}$, ${2\sinh \beta J_2 \simeq e^{\beta J_2}}$ and ${2\cosh \beta J_2 \simeq e^{\beta J_2}}$.
Moreover, in this case, ${2J_z+J_2}$ is equal to ${J_2-|J_x + J_y|}$, and therefore Eq.~(\ref{tripledeg}) reduces to
\begin{align}
\tilde{N} \equiv \frac{ -2 + e^{\beta X  } \tilde{J} /(X + \tilde{J}) }{2 + e^{\beta X } } ,
\end{align}
where $X$ and $\tilde{J}$ are defined  in Eq.~(\ref{JandX}).
From the extremal condition ${d\tilde{N}/dX=0}$, we obtain
\begin{align}
X_{\textrm{op}}\beta &= \log \Bigl( \frac{2\beta X_{\textrm{op}}^2 }{\tilde{J}} + 6\beta  X_{\textrm{op}} +4\beta\tilde{J}    -2   \Bigr) \notag \\
   &\simeq \log 4 \beta \tilde{J},
\end{align}
where we used the same logic as the one with which we derived Eq.~(\ref{Xoptap}) in the doubly degenerate case.
In this way, the optimizing parameter $h_{\textrm{op}}$ and the optimized negativity $N_{\textrm{op}}$ are given as
\begin{align}
h_{\textrm{op}} &\simeq \sqrt{\frac{\tilde{J} }{2\beta} \log4 \beta\tilde{J}  }  ,
\end{align}
and
\begin{align}
N_{\textrm{op}} &\simeq  1- \frac{X_{\textrm{op}}}{ \tilde{J} } - 4e^{-\beta X _{\textrm{op}}} \notag \\
     &\simeq1- \frac{1+ \log 4 \beta \tilde{J} }{\beta \tilde{J} }.
\end{align}
Thus Eq.~(\ref{tripledegenerate}) is proved.
This completes the proof of Theorem~3.

\subsection{Negativity and Concurrence}
We here mention the relationship between the negativity and the concurrence~\cite{Wootter}. The concurrence is also an important entanglement measure. 
Concerning the optimization problem of the concurrence, we can only prove that the negativity $N$ and the concurrence $C$ have the same value for the local parameters ${\{0,0,h,0,0,-h\}}$ with an arbitrary value of $h$; namely,
\begin{align}
N(0,0,h,0,0,-h,\beta) = C(0,0,h,0,0,-h,\beta) . \label{NegaandCon}
\end{align}
This equation is proven by the theorem in Ref.~\cite{Verstraete2}, which says that the concurrence is equal to the negativity iff the eigenvector of $\rho^{T_1}$ corresponding to its negative eigenvalue is a Bell state up to local unitary transformations.

For the local parameters ${\{0,0,h,0,0,-h\}}$, the density matrix ${Z \rho^{T_1}}$ is given in Eq.~\eqref{rhotmat} and only the eigenvalue ${\tilde{N}=a_1- |a_2|}$ can be negative.
For ${\{J_{x},J_{y}\}\ge J_{z}\ge0}$ and ${0\ge J_{z} \ge \{J_{x},J_{y}\}}$ the eigenvectors of $Z\rho^{T_1}$ corresponding to the eigenvalue~${\tilde{N}=a_1- |a_2|}$ is ${(\ket{00} + \ket{11})/\sqrt{2}}$ and ${(\ket{00} - \ket{11})/\sqrt{2}}$, respectively, both being a Bell state.
In the case of ${\tilde{N} >0}$, the concurrence must be equal to the negativity because the eigenvector of $\rho^{T_1}$ corresponding to its negative eigenvalue is a Bell state.
In the case of ${\tilde{N} \le0}$, the negativity ${N=\textrm{max}(\tilde{N},0)}$ is equal to $0$ and the entanglement does not exist. 
Therefore, the concurrence and the negativity are both equal to $0$. This completes the proof of Eq.~\eqref{NegaandCon}

\section{High- and low-temperature phases}
\begin{figure}
\centering
\includegraphics[clip, scale=1]{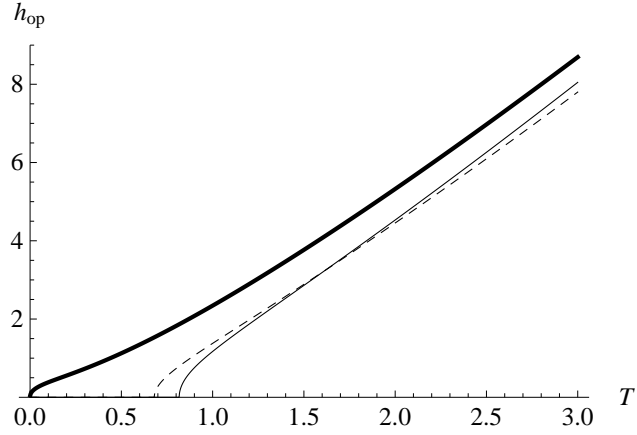}
\caption{Numerically rigorous solution of the optimizing local parameter $h_{\textrm{op}}$: the solid line for ${\{J_{x},J_{y},J_z\} =   \{1/3,1/3,1/3\}}$, the dashed line for ${\{J_{x},J_{y},J_z\} =  \{1/2,1/3,1/6\}}$ and the thick line for ${\{J_{x},J_{y},J_z\} =   \{-1/2,-1/4,-1/4\}}$. The number of data points is $3000$ for each case. The boundary temperatures $T_\textrm{c}$ between the high- and low-temperature phases are $0.8168\cdots$, $0.6803\cdots$ and $0$ for ${\{1/3,1/3,1/3\}}$, ${\{1/2,1/3,1/6\}}$ and ${\{-1/2,-1/4,-1/4\}}$, respectively. 
}
\label{fig:magnetic}
\end{figure}

In the present section, we calculate the optimizing local Hamiltonian and the optimized entanglement numerically rigorously. After the analysis in Sec.~III, we here set $\{h_1^x,h_1^y,h_1^z,h_2^x,h_2^y,h_2^z\}= \{0,0,h_{\textrm{op}},0,0,-h_{\textrm{op}}\}$.
In the calculations below, we will see that there are two kinds of temperature range, which we refer to as the high- and low-temperature phases.  We will find that in the low-temperature phase the optimizing local parameter $h_{\textrm{op}}$ vanishes, whereas in the high-temperature phase it dose not.
We start from Eq.~(\ref{optimizenega}) with the optimizing parameters $\{h_1^x,h_1^y,h_1^z,h_2^x,h_2^y,h_2^z\}= \{0,0,h_{\textrm{op}},0,0,-h_{\textrm{op}}\}$. The parameter $h_{\textrm{op}}$ which maximizes the negativity can be calculated from
\begin{align}
&\frac{1}{h} \frac{\partial \tilde{N}}{ \partial h}\biggl |_{h=h_{\textrm{op}}} \propto |J_x+J_y| ( \beta J_2 \cosh \beta J_2-\sinh  \beta J_2)   \notag \\
&+ J_2^2 \beta \sinh \beta J_2 -\frac{e^{2\beta J_z}|J_x+J_y| }{\cosh \beta J_1}\Bigl(-J_2 \beta +\frac{\sinh 2\beta J_2}{2}\Bigr)  \notag \\
=&0, \label{optimize}
\end{align}
where the factor ${1/h}$ is added to remove the trivial solution of ${h=0}$. In {Fig.~\ref{fig:magnetic}}, we show the optimizing local parameter $h_{\textrm{op}}$ in the cases of ${\{J_{x},J_{y},J_z\} =  \{1/3,1/3,1/3\}}$, ${\{1/2,1/3,1/6\}}$ and ${\{-1/2,-1/4,-1/4\}}$. See Appendix~A for the convergence of $h_{\textrm{op}}$ to the asymptotes~(\ref{happro}) and (\ref{theorem1}). 

\begin{figure}
\centering
\subfigure[]{
\includegraphics[clip, scale=1]{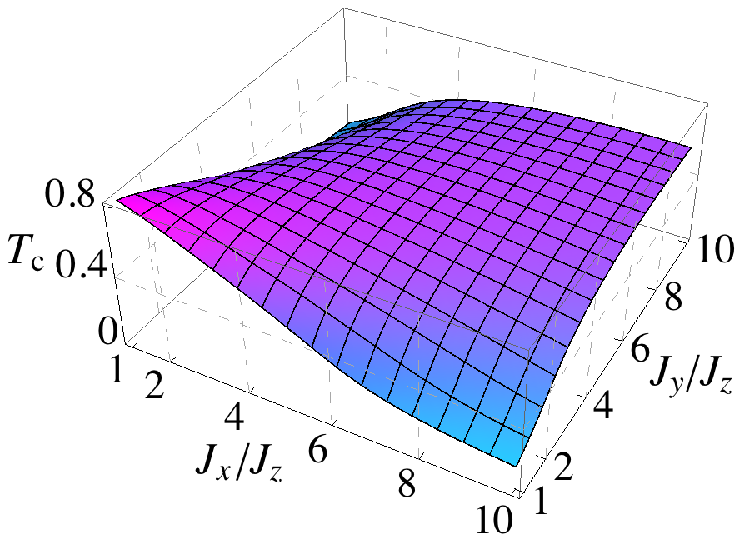}
}
\subfigure[]{
\includegraphics[clip, scale=1]{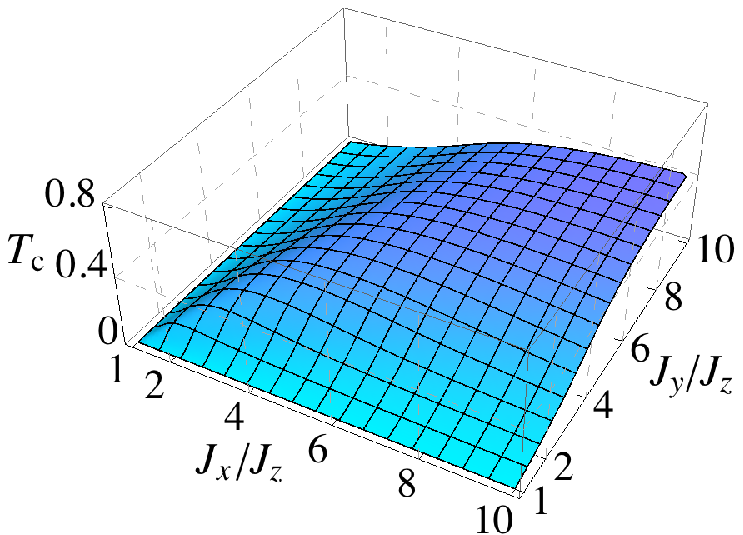}
}
\caption{(color online) The boundary temperature $T_\textrm{c}$ between the high- and low-temperature phases, (a) for the antiferromagnetic case ${\{J_{x},J_{y}\}\ge J_{z}\ge0}$ and (b) for the ferromagnetic case ${0\ge J_{z} \ge \{J_{x},J_{y}\}}$. The point of origin is ${(1,1)}$, which corresponds to the isotropic Heisenberg interaction. In (a), the boundary temperatures are $0.8168\cdots$, $0.1292\cdots$, $0.1292\cdots$, $0.5735\cdots$ and $0.6208\cdots$ at ${(1,1)}$, ${(1,10)}$, ${(10,1)}$, ${(10,10)}$ and ${(5,5)}$, respectively. The maximum temperature is $0.8168\cdots$ at ${(1,1)}$, which is the $XXX$ point.  In (b), the boundary temperatures are $0$, $0$, $0$, $0.4126\cdots$ and $0.3188\cdots$ at ${(1,1)}$, ${(1,10)}$, ${(10,1)}$, ${(10,10)}$ and ${(5,5)}$, respectively. The maximum temperature is $0.5184\cdots$ at ${\lim_{x\rightarrow \infty} {(x,x)}}$, which is the $XX$ point.
}
\label{fig:boundarytemperature}
\end{figure}

In the high-temperature phase, Eq.~(\ref{optimize}) has a non-trivial solution of ${h_{\textrm{op}}>0}$, while in the low-temperature phase, Eq.~(\ref{optimize}) has no solutions and the optimizing value $h_{\textrm{op}}$ is zero, which is the trivial solution of ${\partial \tilde{N}/\partial h=0}$.
Therefore, the boundary temperature $T_\textrm{c}$ between the high- and low-temperature phases is a solution of 
\begin{align}
\lim_{h \rightarrow 0}\frac{1}{h} \frac{\partial \tilde{N}(J_x,J_y,J_z,h,\beta) }{ \partial h} =0. \label{boundary}
\end{align}
The boundary temperature $T_\textrm{c}$ is defined for each interaction Hamiltonian $H_{\textrm{int}}$. 

\begin{figure}
\centering
\subfigure[]{
\includegraphics[clip, scale=1]{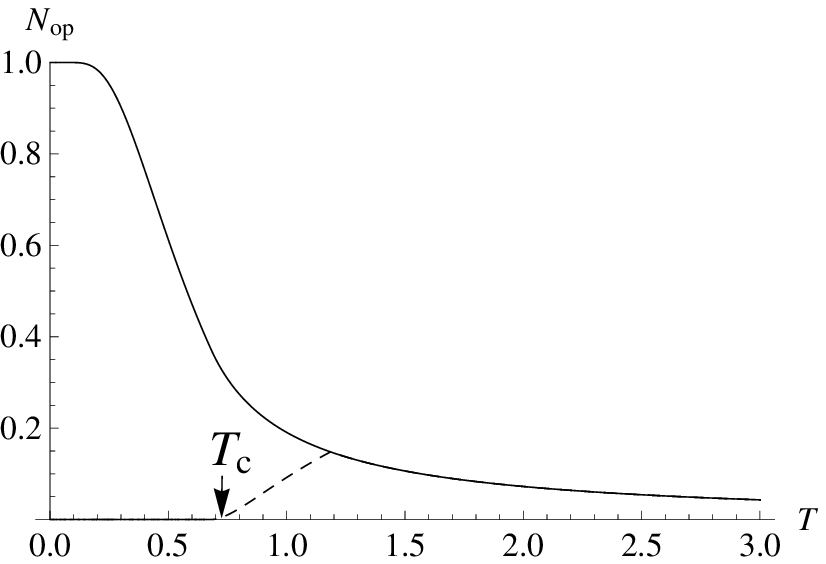}
}
\subfigure[]{
\includegraphics[clip, scale=1]{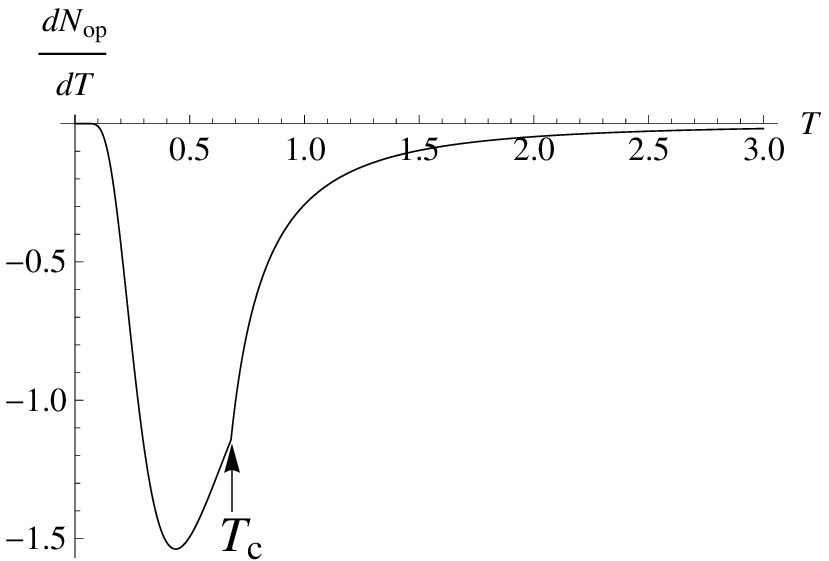}
}
\subfigure[]{
\includegraphics[clip, scale=1]{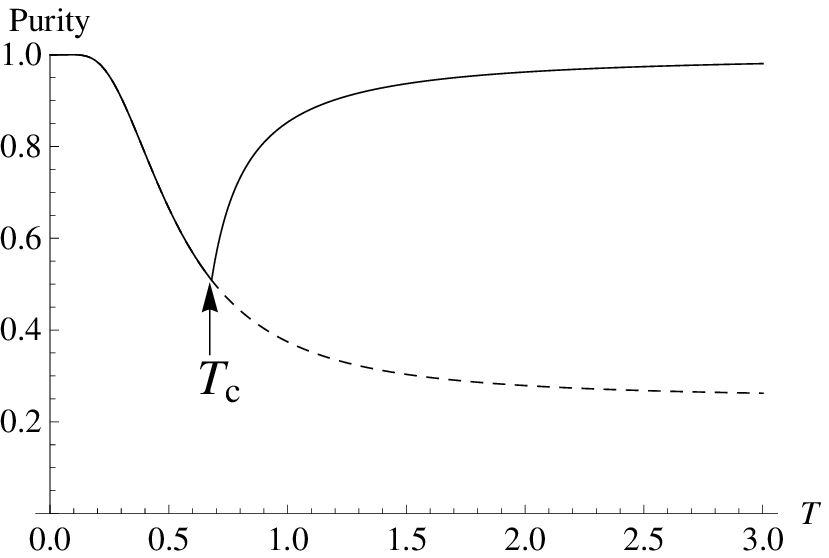}
}
\caption{Plots of (a) the negativity; (b) its first derivatives; (c) the purity, for ${J_x=1/2}$, ${J_y=1/3}$, ${J_z=1/6}$.  The solid line is the optimized entanglement and the dashed line is the entanglement enhancement defined in the text. The boundary temperature is ${T_\textrm{c}=0.6803\cdots}$. At ${T=1.185\cdots}$, the entanglement enhancement is maximum, where the value is $0.1480\cdots$.  In (b), we obtained the data points by the finite-difference method. In (c), the solid line is the purity of the optimized state and the dashed line is the one under no local Hamiltonian. The minimum value of the purity is $0.5087\cdots$ at ${T_\textrm{c}}$.
}
\label{fig:singularity}
\end{figure}

In {Fig.~\ref{fig:boundarytemperature}}, we show the boundary temperature $T_\textrm{c}$ in the cases of ${\{J_{x},J_{y}\}\ge J_{z}\ge0}$ and ${0\ge J_{z} \ge \{J_{x},J_{y}\}}$, which correspond to all kinds of interaction thanks to Lemma~1.
We calculated the data in {Fig.~\ref{fig:boundarytemperature}} from (\ref{boundary}), normalizing the interaction parameters so that ${||H_{\textrm{int}}||_2=1}$, where ${||\ ||_2}$ is the spectral norm. From {Fig.~\ref{fig:boundarytemperature}}, we see the following properties. 
First, the boundary temperatures $T_{\textrm{c}}$ are higher in the antiferromagnetic cases ${\{J_{x},J_{y}\}\ge J_{z}\ge0}$ than in the ferromagnetic cases ${0\ge J_{z} \ge \{J_{x},J_{y}\}}$.
Second, in the antiferromagnetic systems, the boundary temperature $T_{\textrm{c}}$ is maximum of $0.8168\cdots$ for the isotropic Heisenberg interaction (the $XXX$ model).
Next, the boundary temperature $T_{\textrm{c}}$ is zero in the cases of ${0\ge J_{z}=J_{x} \ge J_{y}}$ and ${0 \ge J_{z}=J_{y} \ge J_{x}}$ as well as the case of the ferromagnetic isotropic Heisenberg model, which means that the low-temperature phase shrinks to the zero temperature in these doubly and triply degenerate cases  analyzed in Sec.~III.C.
We have revealed in Sec~III.C that in the low-temperature limit ${\beta \rightarrow \infty}$ the negativity is strictly $1$ with no local Hamiltonian in the non-degenerate cases. The present calculation indeed shows that the low-temperature phase extends to a finite temperature in the non-degenerate cases.
In the antiferromagnetic system, on the other hand, the boundary temperature is zero only in the case of the Ising model, ${J_{x}=J_{z}=0}$ or ${J_{y}=J_{z}=0}$.

Next, we consider the singularity at the boundary between the high- and low-temperature phases.  
In {Fig.~\ref{fig:singularity}}, we show the optimized negativity, its first derivative and the purity ${\textrm{tr}(\rho^2)}$ in the case of ${\{J_{x},J_{y},J_{z}\}=\{1/2, 1/3, 1/6\}}$. We also consider the entanglement enhancement, which is defined as the difference of the entanglement between the optimized entanglement and the entanglement under no local Hamiltonian, namely ${N(H_{\textrm{LO}}^{\textrm{op}})-N({H_{\textrm{LO}}=0})}$.
We numerically rigorously calculated the data in {Fig.~\ref{fig:singularity}(a)} using (\ref{optimize}), and the derivatives by the finite-difference method.
{Figure~\ref{fig:singularity}(b)} shows that the second derivative of the negativity is not continuous at the boundary and {Fig.~\ref{fig:singularity}(c)} shows that the first derivative of the purity is not continuous at the boundary. On the other hand, there is no singularity at the point of ${T=1.185\cdots}$, where the derivative of the entanglement enhancement is not continuous.

The emergence of the high- and low-temperature phases is due to the following reason.
First, the entanglement enhancement by addition of the local Hamiltonian comes from the fact that a local Hamiltonian increases the purity and suppresses the entanglement loss caused by thermal mixing, which is demonstrated in {Fig.~\ref{fig:singularity}(c)}. On the other hand, too strong magnetic fields bring the quantum system close to a direct product states and hence destroy the entanglement. These two effects compete to give rise to the two phases. In the low-temperature phase, we do not need a magnetic field because the purity is already high. In the high-temperature phase, on the other hand, we need a magnetic field because the thermal fluctuation decreases the purity. 
The transition from the low-temperature phase to the high-temperature phase means that the enhancement of the entanglement due to the increase of the purity becomes predominant compared with the entanglement decay caused by the magnetic decoupling.

\section{SUMMARY AND CONCLUSION}\label{conclusion}
We have analytically and numerically rigorously studied thermal states of quantum systems where two qubits interact under a local Hamiltonian $H_{\textrm{LO}}$ and have determined the local Hamiltonian $H_{\textrm{LO}}$ which maximizes the thermal entanglement under a fixed interaction.
As a result, we have found that the interaction Hamiltonian can be transformed into the $XYZ$-exchange interactions whose parameters are either antiferromagnetic as ${\{J_{x},J_{y}\}\ge J_{z}\ge0}$ or ferromagnetic as ${0\ge J_{z} \ge \{J_{x},J_{y}\}}$ and that the optimizing local Hamiltonian always takes the form of ${h_{\textrm{op}}(\sigma_1^z\otimes I -I \otimes \sigma_2^z)}$, where $h_{\textrm{op}}$ depends on the temperature. In addition, we have proved that the optimized entanglement does not vanish at any temperatures and decays slowly according to ${1/(T\log T)}$ at high temperatures. We have also found that in the low-temperature phase the entanglement is maximum without any local Hamiltonian and have investigated the interaction dependence of the boundary temperature of this range. Indeed, the low-temperature phase shrinks to the zero temperature point if the interaction Hamiltonian has degeneracy. At the same time, we have discovered a singularity of the optimized entanglement at the boundary temperature, where the second derivative is discontinuous. 

In conclusion, our work has revealed general properties of the thermal entanglement of interacting two qubits, though we have assumed a numerically confirmed hypothesis. The concept of high- and low-temperature phases is an interesting property in that it is based on the response to external manipulation of local Hamiltonians. It is likely that we can find more interesting properties of entanglement in this regard. In future, we plan to investigate two qubits which interact indirectly or general bipartite systems.
\section*{ACKNOWLEDGMENT}
We are grateful to Dr.~H.~Azuma for critical reading.
The present study is supported by CREST from Japan Science and Technology Agency as well as Grant-in-Aid for scientific Research No.~22340110.
\appendix
\section{Numerical comparison}\label{apendixA}

\begin{figure}
\centering
\subfigure[]{
\includegraphics[clip, scale=1]{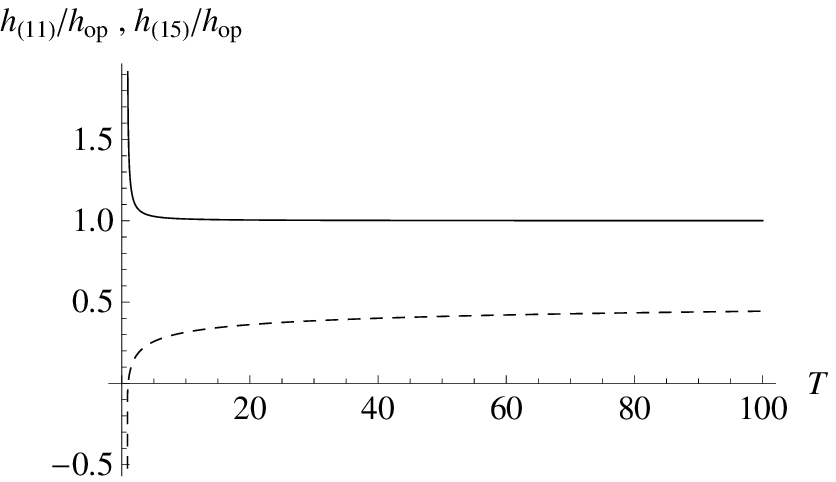}
}
\subfigure[]{
\includegraphics[clip, scale=]{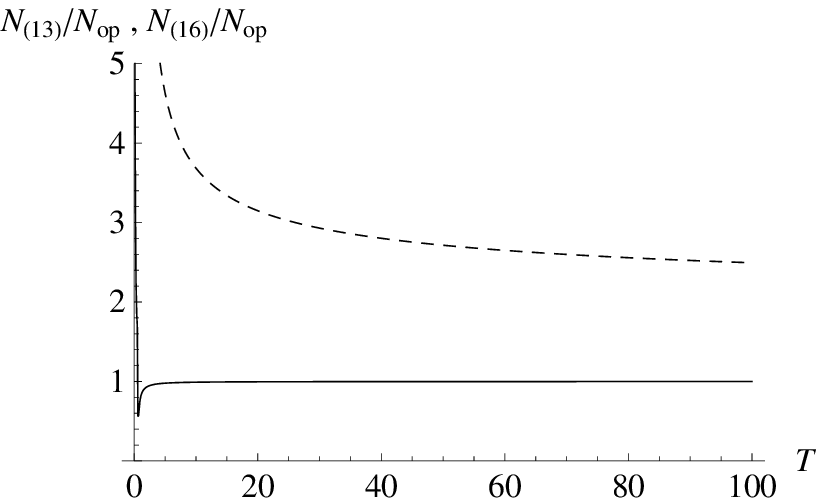}
}
\caption{The comparison between the asymptotes in {Eqs.~(\ref{happro})} and {(\ref{Nappro})} and those in {Eqs.~(\ref{theorem1})} and {(\ref{theorem1N})}.
 (a) for the ratios $h_{(\ref{happro})}/h_{\textrm{op}}$ (solid line) and ${h_{(\ref{theorem1})}/h_{\textrm{op}}}$ (dashed line), where $h_{(\ref{happro})}$ and $h_{(\ref{theorem1})}$ are derived from Eqs.~(\ref{appcompa}) and (\ref{appcompc}), respectively, and $h_{\textrm{op}}$ is the numerically rigorous value calculated from Eq.~(\ref{optimize}). At ${T=100}$, the values of $h_{\textrm{op}}^{(\ref{happro})}/h_{\textrm{op}}$ and $h_{\textrm{op}}^{(\ref{theorem1})}/h_{\textrm{op}}$ are $1.0007$ and $0.4438$, respectively.
(b) for the ratios $N_{(\ref{Nappro})}/N_{\textrm{op}}$ (solid line) and $N_{(\ref{theorem1N})}/N_{\textrm{op}}$ (dashed line), where $N_{(\ref{Nappro})}$ and $N_{(\ref{theorem1N})}$ are derived from Eqs.~(\ref{appcompb}) and (\ref{appcompd}), respectively, and $N_{\textrm{op}}$ is the numerically rigorous value calculated from Eq.~(\ref{optimize}). At ${T=100}$, the values of $N_{(\ref{Nappro})}/N_{\textrm{op}}$ and $N_{(\ref{theorem1N})}/N_{\textrm{op}}$ are $0.9994$ and $2.494$, respectively.
}
\label{fig:Approximation}
\end{figure}

In the present Appendix, we compare the asymptotes in {Eqs.~(\ref{happro})} and {(\ref{Nappro})} with those in {Eqs.~(\ref{theorem1})} and {(\ref{theorem1N})}
in the case of ${\{J_{x},J_{y},J_z\} =   \{1/3,1/3,1/3\}}$. In this case, {Eqs.~(\ref{happro})}, {(\ref{Nappro})}, {(\ref{theorem1})} and {(\ref{theorem1N})}, respectively, reduce to\begin{align}
e^{2h'_{(\ref{happro})}} &= \frac{ 12 h'^2_{(\ref{happro})} } {\beta}, \label{appcompa} \\  
N_{(\ref{Nappro})} &=\beta \frac{1}{3h'_{(\ref{happro})}} - 2e^{-2h'_{(\ref{happro})}}, \label{appcompb} \\ 
h_{(\ref{theorem1})}&= \frac{\log 1/\beta}{2\beta}, \label{appcompc} \\    
N_{(\ref{theorem1N})} &= \beta \frac{2}{3\log 1/\beta},  \label{appcompd}
\end{align}
where the subscripts denote the equation number of the corresponding asymptotes. 
In {Fig.~\ref{fig:Approximation}}, we show the comparison of these asymptotes with the numerically rigorous estimates of $h_{\textrm{op}}$ and $N_{\textrm{op}}$ obtained from Eq.~(\ref{optimize}).
We can see that the convergences of ${h_{(\ref{theorem1})}/h_{\textrm{op}}}$ and $N_{(\ref{theorem1N})}/N_{\textrm{op}}$ are very slow, while the convergences of $h_{(\ref{happro})}/h_{\textrm{op}}$ and $N_{(\ref{Nappro})}/N_{\textrm{op}}$ are much faster. The convergence of $N(h_{(\ref{happro})})/N_{\textrm{op}}$, where $N(h)$ is given in Eqs.~(\ref{optnega}) and (\ref{optimizenega}), is even faster than that of $N_{(\ref{Nappro})}/N_{\textrm{op}}$; at $T=100$, the values of $N(h_{(\ref{happro})})/N_{\textrm{op}}$ and $N_{(\ref{Nappro})}/N_{\textrm{op}}$ are $0.999998$ and $0.9994$, respectively.

\section{Lemma~2 in degenerate cases}\label{apendixB}
In the proof of Lemma~2, we left out the cases of ${\zeta=0}$ or ${\zeta=\pm 1}$ in Eq.~(\ref{Zrho}). In the present Appendix, we prove that Lemma~2 still holds in these cases. First, the general form of Eq.~(\ref{perturbation}) is given in the basis of $\{\ket{\mu}\}_{\mu=1}^4$ as follows:
\begin{align}
&Z\rho^{T_1}  \notag \\
&\xrightarrow{\beta \rightarrow 0} \begin{pmatrix}  e^{2h'} & a_{12} f_{12} \beta & a_{31}f_{31}  \beta&a_{32}f_{32}\beta \\ a_{21}f_{21}\beta  & e^{2\zeta h'} &a_{41}f_{41}\beta& a_{42} f_{42}\beta \\ a_{13}f_{13}\beta& a_{14}f_{14}\beta &  e^{-2\zeta h'} &  a_{34}f_{34}\beta\\ a_{23}f_{23}\beta&a_{44} f_{42}\beta&a_{43}f_{43} \beta& e^{-2 h'}  \end{pmatrix}, \label{genZrho}
\end{align}
where $\{a_{ij}\}$ are constants of order 1 and $\{f_{\mu \nu}\}$ are defined in Eq.~(\ref{fij}). Note that on the diagonal of Eq.~\eqref{genZrho}, the second term of Eq.~(\ref{perturbation}) is neglected in comparison to the first term. In the cases of $\zeta \neq 0$ and $\zeta \neq \pm 1$, Eq.~(\ref{genZrho}) reduces to Eq.~(\ref{Zrho}).

In the case of ${\zeta = 0}$, we have $\{E'_1,E'_2,E'_3,E'_4\}=\{2h',0,0,-2h'\}$ and ${f_{23}=f_{32}=e^{-E'_2}=1}$, and hence Eq.~(\ref{genZrho}) reduces to
\begin{align}
&Z\rho^{T_1}  \notag \\
&\xrightarrow{\beta \rightarrow 0} \begin{pmatrix}  e^{2h'} & a_{12}\frac{\beta e^{2h'}}{h'} & a_{31}\frac{\beta e^{2h'}}{h'} & a_{32}\beta  \\ a_{21}\frac{\beta e^{2h'}}{h'}  & 1 &a_{41}\frac{\beta e^{2h'}}{h'} &a_{42} \frac{\beta}{h'}    \\ a_{13}\frac{\beta e^{2h'}}{h'}  &a_{14}\frac{\beta e^{2h'}}{h'}  &  1 &  a_{34}\frac{\beta }{h'}  \\ a_{23} \beta & a_{24} \frac{\beta}{h'} &a_{43}\frac{\beta}{h'} & e^{-2 h'}  \end{pmatrix}, 
\end{align} 
In this case, the product of the diagonal elements (PD) of ${Z\rho^{T_1}}$ is $1$, whereas the maximum of the absolute values of the products including off-diagonal elements (POD) is of order ${e^{4h'} \beta^2/h'^2}$, which comes from the product $ {-e^{2h'}\times {a_{41}\frac{\beta e^{2h'}}{h'}\times a_{14}\frac{\beta e^{2h'}}{h'}} \times e^{-2h'}}$. Therefore, it is necessary for ${\det \rho^{T_1}<0}$ that the order of ${e^{4h'} \beta^2/h'^2}$ is greater or of order $1$, which leads to 
\begin{align}
\beta h =h' > \frac{\log 1/\beta}{2} \label{betah}
\end{align}
as in Eq.~\eqref{hlower}.
Thus, Lemma~2 is proved in the case of ${\zeta = 0}$.

The proofs for the cases of ${\zeta = 1}$ and ${\zeta = -1}$, or the cases of ${\{h_1, h_2\} =\{2h,0\}}$ and ${\{h_1, h_2\} =\{0,2h\}}$, are essentially the same. We here present the proof only for the case of ${\zeta = 1}$. In this case, we have $\{E'_1,E'_2,E'_3,E'_4\}=\{2h',2h',-2h',-2h'\}$, ${f_{12}=f_{21}=e^{2h'}}$ and ${f_{34}=f_{43}=e^{-2h'}}$, and hence Eq.~(\ref{genZrho}) reduces to
\begin{align}
&Z\rho^{T_1}  \notag \\
&\xrightarrow{\beta \rightarrow 0} \begin{pmatrix}  e^{2h'} & a_{12} \beta e^{2h'}  & a_{31}\frac{\beta e^{2h'}}{h'}  &a_{32}\frac{\beta e^{2h'}}{h'} \\ a_{21} \beta e^{2h'}  & e^{2  h'} &a_{41}\frac{\beta e^{2 h'}}{h'}& a_{42} \frac{\beta e^{2  h'}}{h'} \\ a_{13}\frac{\beta e^{2h'}}{h'} & a_{14}\frac{\beta e^{2  h'}}{h'} &  e^{-2  h'} &  a_{34} \beta e^{-2 h'}  \\ a_{23}\frac{\beta e^{2h'}}{h'} &a_{24} \frac{\beta e^{2 h'}}{h'} &a_{43} \beta e^{-2 h'}  & e^{-2 h'}  \end{pmatrix}, 
\end{align}
The PD of ${Z\rho^{T_1}}$ is 1, whereas the maximum of the absolute values of the PODs is of order ${e^{4h'} \beta^2/h'^2}$ or of order ${e^{8h'} \beta^4/h'^4}$, which come from $ {-e^{2h'} \times {a_{41}\frac{\beta e^{2h'}}{h'} \times a_{14}\frac{\beta e^{2h'}}{h'}}\times  e^{-2h'}}$ and ${a_{32} \frac{\beta e^{2h'}}{h'} \times a_{41} \frac{\beta e^{2h'}}{h'} \times a_{14} \frac{\beta e^{2h'}}{h'}  \times a_{23}\frac{\beta e^{2h'}}{h'}}$, respectively. Therefore, it is also necessary for ${\det \rho^{T_1}<0}$ that ${e^{4h'} \beta^2/h'^2}$ is greater or of order 1, which again leads to Eq.~(\ref{betah}). Thus, Lemma~2 is also proved in the case of ${\zeta = 1}$. 

\section{Proof of  Eq.~(\ref{taisyou})}%\label{apendixC}
In order to prove Eq.~(\ref{taisyou}), we begin with the standard operator expansion of an arbitrary ${2\otimes2}$ operator $Q$:
\begin{align}
Q=\frac{1}{4} \sum_{i,j=0,x,y,z} q_{ij} \sigma_1^i \otimes \sigma _2^j,
\end{align}
where ${\sigma_1^0=\sigma_2^0=I}$ is the two-dimensional identity operator. The coefficients $q_{ij}$ are given by
\begin{align}
q_{ij} = \textrm{tr} \bigl(Q \sigma_1^i \otimes \sigma _2^j \bigr)
\end{align}
because ${\textrm{tr}(I \otimes I)=4}$ and the other terms are traceless.

Symmetries that the Hamiltonian~\eqref{Htotop}
%\begin{align}\label{bessi1-30}
%H_\mathrm{tot}^\mathrm{op}=\sum_{i=x,y,z}J_i\sigma_1^i\otimes\sigma_2^i+h(\sigma_1^z\otimes I-I\otimes\sigma_2^z)
%\end{align}
possesses eliminate many of the coefficients $\{q_{ij}\}$ of the expansion of operators with the same symmetries, such as $\exp(-\beta x H_\mathrm{tot}^\mathrm{op})$.
First, a straightforward calculation shows that the Hamiltonian~\eqref{Htotop} commutes with the global phase flip
\begin{align}
U_\mathrm{flip}&=e^{i(\pi/2)\sigma_1^z}\otimes e^{i(\pi/2)\sigma_2^z}
\nonumber\\
&=-\sigma_1^z\otimes\sigma_2^z.
\end{align}
This operator flips the signs of $\sigma^x$ and $\sigma^y$.
For an operator $Q$ that commutes with $U_\mathrm{flip}$, the coefficients ${\{q_{0x},q_{0y},q_{x0},q_{y0},q_{xz},q_{yz},q_{zx},q_{zy}\}}$ vanish.
For example, we have
\begin{align}
q_{xz}&=\mathrm{tr}[Q(\sigma_1^x\otimes\sigma_2^z)]
\nonumber\\
&=\mathrm{tr}[{U_\mathrm{flip}}^{-1}QU_\mathrm{flip}{U_\mathrm{flip}}^{-1}(\sigma_1^x\otimes\sigma_2^z)U_\mathrm{flip}]
\nonumber\\
&=\mathrm{tr}[Q((-\sigma_1^x)\otimes\sigma_2^z)]
\nonumber\\
&=-q_{xz}=0.
\end{align}
The same argument gives $q_{0x}=q_{0y}=q_{x0}=q_{y0}=q_{xz}=q_{yz}=q_{zx}=q_{zy}=0$.

Next, the Hamiltonian~\eqref{Htotop} is a real matrix in the $\sigma^z$ basis.
Noting that only $\sigma^y$ has imaginary elements in this representation, we have, for an operator $Q$ with the symmetry ${Q^\ast=Q}$, 
\begin{align}
(q_{xy})^\ast&=\mathrm{tr}[Q^\ast ((\sigma_1^x)^\ast\otimes(\sigma_2^y)^\ast)]
\nonumber\\
&=\mathrm{tr}[Q (\sigma_1^x\otimes(-\sigma_2^y))]
\nonumber\\
&=-q_{xy}.
\end{align}
On the other hand, the Hermiticity of an operator $Q$ is followed by
\begin{align}
(q_{xy})^\ast&=\mathrm{tr}[((\sigma_1^x)^\dag\otimes(\sigma_2^y)^\dag) Q^\dag]
\nonumber\\
&=\mathrm{tr}[Q (\sigma_1^x\otimes\sigma_2^y)]
\nonumber\\
&=q_{xy}.
\end{align}
The above argument shows $q_{xy}=q_{yx}=0$.

Finally, the Hamiltonian~\eqref{Htotop} is symmetric with respect to the following set of operations:
\begin{align}
U_{12}&=(e^{i(\pi/2)\sigma_1^x}\otimes e^{i(\pi/2)\sigma_2^x})P_{12}
\nonumber\\
&=-(\sigma_1^x\otimes\sigma_2^x)P_{12},
\end{align}
where $P_{12}$ is the permutation of the spins 1 and 2.
The operator ${\sigma_1^x\otimes\sigma_2^x}$ flips the signs of $\sigma_1^z$ and $\sigma_2^z$ but the permutation $P_{12}$ makes the signs back to the original ones, because the local fields are in the opposite directions in the Hamiltonian~\eqref{Htotop}.
For an operator $Q$ that commutes with $U_{12}$, we have
\begin{align}
q_{z0}&=\mathrm{tr}[Q(\sigma_1^z\otimes I)]
\nonumber\\
&=\mathrm{tr}[{U_{12}}^{-1}QU_{12}{U_{12}}^{-1}(\sigma_1^z\otimes I)U_{12}]
\nonumber\\
&=\mathrm{tr}[Q(I\otimes(-\sigma_2^z))]
\nonumber\\
&=-q_{0z}.
\end{align}

To summarize, an operator with the same symmetries as the Hamiltonian~\eqref{Htotop} is expanded in the form
\begin{align}\label{bessi-eq100}
Q&=\frac{1}{4}\Biggl[q_{00}I\otimes I+q_{z0}(\sigma_1^z\otimes I - I\otimes\sigma_2^z)
\nonumber\\
&+\sum_{i=x,y,z}q_{ii}\sigma_1^i\otimes\sigma_2^i\Biggr].
\end{align}
In \eqref{setudoukou}, the operators $e^{-\beta x H_\mathrm{tot}^\mathrm{op}}$ and $e^{-\beta (1-x) H_\mathrm{tot}^\mathrm{op}}$ have the same symmetries as the Hamiltonian $H_\mathrm{tot}^\mathrm{op}$ and hence are given in the form~\eqref{bessi-eq100}.

Since the density operator ${\rho=e^{-\beta H_\mathrm{tot}^\mathrm{op}}}$ is given in the form~\eqref{bessi-eq100}, the partial transpose $\rho^{T_1}$ is also of the form~\eqref{bessi-eq100};
in the $\sigma_z$ basis, the partial transpose $T_1$ only flips the sign of $\sigma_1^y$ and hence changes only the sign of $q_{yy}$ in the expansion, not the symmetries nor the form of the expansion.

The state $|\phi_-\rangle$ is a non-degenerate eigenstate of the operator $\rho^{T_1}$ if the minimum eigenvalue $\lambda_-$ is negative.
Suppose that the operator $\rho^{T_1}$ commutes with a symmetry operator $U$.
Then the projection operator $|\phi_-\rangle\langle\phi_-|$ should have the same symmetry.
This is shown as follows. Since we have
\begin{align}
\rho^{T_1}U|\phi_-\rangle=U\rho^{T_1}|\phi_-\rangle=\lambda_-U|\phi_-\rangle
\end{align}
and $|\phi_-\rangle$ is non-degenerate, the vector $U|\phi_-\rangle$ must be the same vector as $|\phi_-\rangle$ except for a phase: ${U|\phi_-\rangle=e^{i\theta}|\phi_-\rangle}$.
Therefore, the projection operator $|\phi_-\rangle\langle\phi_-|$ commutes with $U$ if the negativity is non-zero.
This means that $|\phi_-\rangle\langle\phi_-|$ as well as $(|\phi_-\rangle\langle\phi_-|)^{T_1}$ have the same symmetries as the Hamiltonian $H_\mathrm{tot}^\mathrm{op}$ and are expanded in the form~\eqref{bessi-eq100}.

We thereby arrive at the conclusion that the operator
\begin{align}
&e^{- \beta x H_{\textrm{tot}}^{\textrm{op}} }  \hat{n} e^{- \beta (1-x) H_{\textrm{tot}}^{\textrm{op}} }  \notag \\
=&e^{- \beta x H_{\textrm{tot}}^{\textrm{op}} }  \Bigl[ N(\rho_{\textrm{op}}) \bigl(I\otimes I \bigr)  +2 \bigl(\ket{ \phi_- } \bra{ \phi_- }\bigr)^{T_1}\Bigr] e^{- \beta (1-x) H_{\textrm{tot}}^{\textrm{op}} }  
\end{align}
has the same symmetries as the Hamiltonian $H_\mathrm{tot}^\mathrm{op}$ and hence is expanded in the form~\eqref{bessi-eq100}.

\section{The Eigenvalues of (\ref{rhotmat})}%\label{apendixC}
In this section, we prove that in the eigenvalues of the matrix~(\ref{rhotmat}), only ${a_1-|a_2|}$ can have a negative value for ${\{J_{x},J_{y}\}\ge J_{z}\ge0}$ and ${0\ge J_{z} \ge \{J_{x},J_{y}\}}$.
The four eigenvalues are given in~\eqref{rhoeigenvalues}. Because ${a_1>0}$, ${|a_2|>0}$ and ${b_2^2+b_3^2>0}$, we obviously have
\begin{align}
a_1+|a_2|>0,\ \ b_1+\sqrt{b_2^2+b_3^2}>0.
\end{align}
Therefore, we only have to prove that ${b_1-\sqrt{b_2^2+b_3^2}>0}$.

First, we prove this inequality for ${h=0}$.
For ${h=0}$, the eigenvalue ${b_1-\sqrt{b_2^2+b_3^2}}$ reduces to
\begin{align}
&b_1-\sqrt{b_2^2+b_3^2}  \notag \\
&= e^{\beta J_z} \cosh \bigl[ \beta (J_x + J_y)\bigr]- e^{-\beta J_z} \sinh \bigl[\beta |J_x - J_y|\bigr] \notag \\
&=\frac{1}{2} \bigl( e^{\beta (J_x + J_y +J_z)} + e^{\beta (-J_x - J_y +J_z)} \notag \\
            &\ \ \ \  - e^{\beta (|J_x - J_y|-J_z)}+e^{\beta (-|J_x - J_y|-J_z)}  \bigr).
\end{align}
For ${\{J_{x},J_{y}\}\ge J_{z}\ge0}$, we have 
\begin{align}
e^{\beta (J_x + J_y +J_z)} - e^{\beta (|J_x - J_y|-J_z)}\ge0,
\end{align}
which leads to ${b_1-\sqrt{b_2^2+b_3^2}>0}$.
For ${0\ge J_{z} \ge \{J_{x},J_{y}\}}$, we have 
\begin{align}
&e^{\beta (-J_x - J_y +J_z)}  - e^{\beta (|J_x - J_y|-J_z)} \notag \\
=& \begin{cases}
                           2e^{-\beta J_y} \sinh \bigl[\beta (-J_x +J_z) \bigr] &\textrm{for}\  0\ge J_{z} \ge J_{x}\ge J_{y},  \\
                         2 e^{-\beta J_x} \sinh \bigl[\beta (-J_y +J_z) \bigr] &\textrm{for}\  0\ge J_{z} \ge J_{y}\ge J_{x}.
                            \end{cases}  
\end{align} 
Because ${-J_x +J_z\ge0}$ and ${-J_y +J_z\ge0}$,
\begin{align}
e^{\beta (-J_x - J_y +J_z)} - e^{\beta (|J_x - J_y|-J_z)}\ge0
\end{align}
for ${0\ge J_{z} \ge \{J_{x},J_{y}\}}$, which also leads to ${b_1-\sqrt{b_2^2+b_3^2}>0}$.
 Thus, ${b_1-\sqrt{b_2^2+b_3^2}>0}$ is proved for ${h=0}$.

Next, we prove ${b_1^2- b_2^2 -b_3^2>0}$ for arbitrary $h$, which is equivalent to ${b_1-\sqrt{b_2^2+b_3^2}>0}$ because ${b_1+\sqrt{b_2^2+b_3^2}>0}$.
The value of ${b_1^2- b_2^2 -b_3^2}$ is calculated as follows:
\begin{align}
&b_1^2- b_2^2 -b_3^2 \notag \\
=&e^{2\beta J_z} \biggl( \cosh^2 \beta J_2-\frac{4h^2}{J_2^2}\sinh^2 \beta J_2 \biggr)-e^{-2\beta J_z}\sinh^2 \beta J_1 \notag \\
=&e^{2\beta J_z} \biggl[1+ \Bigl( 1- \frac{4h^2}{J_2^2}\Bigr) \sinh^2 \beta J_2 \biggr]-e^{-2\beta J_z}\sinh^2 \beta J_1 \notag \\
=&e^{2\beta J_z} +\frac{e^{2\beta J_z} (J_x+J_y)^2}{J_2^2}\sinh^2 \beta J_2 -e^{-2\beta J_z}\sinh^2 \beta J_1.
\end{align} 
Only the second term depends on $h$ through ${J_2=\sqrt{4h^2 + (J_x+J_y)^2}}$. The term $(\sinh \beta J_2/J_2)^2$ is a monotonically increasing function of $J_2$ for ${J_2>0}$, while $J_2$ is a monotonically increasing function of $h^2$.
Therefore, ${b_1^2- b_2^2 -b_3^2}$ is also a monotonically increasing function of $h^2$.
Since we already proved that ${b_1^2- b_2^2 -b_3^2}$ is positive for ${h=0}$, we obtain ${b_1^2- b_2^2 -b_3^2>0}$ for any values of $h$, and thus ${b_1-\sqrt{b_2^2+b_3^2}>0}$ is proved.

\renewcommand{\refname}{\vspace{-1cm}}


\begin{thebibliography}{00}
\bibitem{Nielsen} 
M. A. Nielsen and I. L. Chuang, \textit{Quantum computation and Quantum information} (Cambridge University Press, Cambridge, 2000).

\bibitem{horodecki} 
R. Horodecki, P. Horodecki, M. Horodecki, and K. Horodecki, Rev. Mod. Phys. {\bf 81}, 865 (2009).

\bibitem{Amico} 
L. Amico, R. Fazio, A. Osterloh, and V. Vedral, Rev. Mod. Phys. {\bf 80}, 517 (2008).

\bibitem{Arnesen} 
M. C. Arnesen, S. Bose, and V. Vedral, Phys. Rev. Lett. {\bf 87}, 017901 (2001).

\bibitem{Abliz} 
A. Abliz, H. J. Gao, X. C. Xie, Y. S. Wu, and W. M. Liu, Phys. Rev. A {\bf 74}, 052105 (2006). 

\bibitem{Romano} 
R. Romano and D. D'Alessandro, Phys. Rev. Lett. {\bf 97}, 080402 (2006).

\bibitem{Manicini} 
S. Mancini and H. M. Wiseman, Phys. Rev. A {\bf 75}, 012330 (2007).

\bibitem{Plastina} 
F. Plastina and T. J. G. Apollaro, Phys. Rev. Lett. {\bf 99}, 177210 (2007).

\bibitem{Sabrina} 
S. Maniscalco, F. Francica, R. L. Zaffino, N. Lo Gullo, and F. Plastina, Phys. Rev. Lett. {\bf 100}, 090503 (2008).

\bibitem{Kamta} 
G. L. Kamta and A. F. Starace, Phys. Rev. Lett. {\bf 88}, 107901 (2002).

\bibitem{Sun} 
Y. Sun, Y. Chen, and H. Chen, Phys. Rev. A {\bf 68}, 044301 (2003).

\bibitem{Chan} 
W. L. Chan, D. Yang, and S. J. Gu, Chin. Phys. Lett. {\bf 25}, 832 (2008).

\bibitem{Gurkan} 
Z. N. Gurkan and O. K. Pashaev, quant-ph/0804.0710v2. 

\bibitem{Fardin} 
F. Kheirandish, S. J. Akhtarshenas, and H. Mohammadi, Phys. Rev. A {\bf 77}, 042309 (2008).

\bibitem{Bose} 
D. Gunlycke, V. M. Kendon, V.Vedral, and S. Bose, Phys. Rev. A {\bf 64}, 042302 (2001).


\bibitem{XWang}
X. Wang, Phys. Lett. A {\bf 281}, 101 (2001).


\bibitem{Asoudeh} 
M. Asoudeh and V. Karimipour, Phys. Rev. A {\bf 71}, 022308 (2005).

\bibitem{GFZhang} 
G. F. Zhang and S. S. Li, Phys. Rev. A {\bf 72}, 034302 (2005).

\bibitem{Wootter} 
W. K. Wootters, Phys. Rev. Lett. {\bf 80}, 2245 (1998).

\bibitem{Vidal} 
G. Vidal and R. F. Werner, Phys. Rev. A {\bf 65}, 032314 (2002).

\bibitem{Augusiak}  
R. Augusiak, M. Demianowicz, and P. Horodecki, Phys. Rev. A {\bf 77}, 030301(R) (2008)


\bibitem{Dzyaloshinskii}
 I. Dzyaloshinskii, J. Phys. Chem. Solids  {\bf 4}, 228 (1958)
 
\bibitem{Moriya} 
T. Moriya, Phys. Rev. {\bf 117}, 635 (1960).

\bibitem{Matrix Analysis} 
R. Horn and C. Johnson, \textit{Matrix Analysis} (Cambridge University Press, Cambridge, 1985).

\bibitem{Verstraete1} 
F. Verstraete, K. Audenaert, and  B. De Moor, Phys. Rev. A {\bf 64}, 012316 (2001).

\bibitem{Peres} 
A. Peres, Phys. Rev. Lett {\bf 77}, 1413 (1996).

\bibitem{Horodecki} 
M. Horodecki, P. Horodecki, and R. Horodecki, Phys. Lett A {\bf 223}, 1 (1996).

%\bibitem{Miranowicz} %concurrence negativity
%G. Vidal and R. Tarrach, Phys. Rev. A {\bf 59}, 141 (1999).

%\bibitem{Moor} %concurrence=negativity
%K. Audenaert, F. Verstraete, T. D. Bie and B. De. Moor, quant-ph/0012074. 

\bibitem{Wfunction} 
R. M. Corless, G. H. Gonnet, D. E. G. Hare, D. J. Jeffrey and D. E. Knuth, Adv. Comput. Math. {\bf 5}, 329 (1996).


\bibitem{Osborne} 
T. J. Osborne and M. A. Nielsen, Phys. Rev. A {\bf 66}, 032110 (2002).

\bibitem{Verstraete2} 
F. Verstraete, K. Audenaert, J. Dehaene, and  B. De Moor, J. Phys. A: Math. Gen. {\bf 34}, 10327 (2001).






% 
% 
% 
\end{thebibliography}
\end{document}